\documentclass[11pt]{article}

\usepackage[a4paper,left=2.4cm,right=2.4cm,top=2.7cm,bottom=2.8cm]{geometry}

\usepackage[numbers,sort&compress]{natbib}
\usepackage[T1]{fontenc}
\usepackage{newtxtext,newtxmath}
\usepackage{microtype}
\usepackage{physics}
\usepackage{amsmath}
\numberwithin{equation}{section}

\usepackage{graphicx}
\usepackage{subcaption}
\usepackage{booktabs}

\usepackage{authblk}

\usepackage{xcolor}
\usepackage[colorlinks=true,
            linkcolor=blue!60!black,
            citecolor=blue!60!black,
            urlcolor=blue!60!black]{hyperref}

\usepackage{titlesec}
\titleformat{\section}
  {\large\bfseries}
  {\thesection.}{0.5em}{}

\titleformat{\subsection}
  {\normalsize\bfseries}
  {\thesubsection.}{0.5em}{}

\title{\bfseries Monotonic probes of asymmetry and robustness of the quantum Mpemba effect}

\author[1]{Riccardo Travaglino}
\author[2]{Michele Mazzoni}
\author[1]{Pasquale Calabrese}
\affil[1]{SISSA and INFN Sezione di Trieste, Via Bonomea 265, I-34136 Trieste, Italy}
\affil[2]{Dipartimento di Fisica e Astronomia, Università di Bologna and INFN, Sezione di Bologna, Italy}

\date{}

\begin{document}

\maketitle

\begin{abstract}
Entanglement asymmetry has emerged as a powerful probe of symmetry restoration and the quantum Mpemba effect. However, its conventional Rényi generalizations, defined as differences of Rényi entropies, are not monotones in the resource-theoretic sense. Here, we investigate an alternative family of asymmetry quantifiers based on sandwiched Rényi divergences, which satisfy monotonicity under symmetry-preserving quantum channels. Developing a two-parameter replica approach, we show how these quantities can be computed from charged moments and establish a remarkable correspondence with conventional Rényi asymmetries under the inversion of the Rényi index, $\alpha\leftrightarrow 1/\alpha$. This relation, exact for pure states, holds at leading order in the relevant scaling limits for homogeneous quantum quenches, including interacting integrable systems, and for Haar-random states. Applying these results to quenches from tilted ferromagnetic states, we demonstrate that the quantum Mpemba effect is remarkably robust, occurring throughout the entire family of sandwiched Rényi asymmetry monotones. We identify dynamical regimes in quenches from spatially inhomogeneous initial states where conventional Rényi asymmetries and genuine monotones exhibit qualitatively different behaviours. We also investigate a solvable toy model illustrating the possible dependence of Mpemba crossings on the Rényi index. Our findings establish a connection between conventional Rényi asymmetries and quantum resource theory, explaining their effectiveness in broad classes of nonequilibrium protocols while highlighting the additional information provided by genuine asymmetry monotones.
\end{abstract}

\section{Introduction} \label{sec:intro} Recent years have witnessed a surge of interest in the interplay between quantum many-body dynamics and quantum information theory. While conventional quantum many-body theory has primarily focused on local observables and correlation functions, shifting attention to probes that reveal finer properties of quantum states has substantially enriched our understanding of both low-energy and non-equilibrium physics. Foremost among such quantities is entanglement \cite{Plenio2005AnIT,damico2008,Calabrese_2009,horodecki2009}, which has proven to be a central ingredient in characterizing the ground states of local Hamiltonians \cite{vidal_2003,Pasquale_Calabrese_2004,Laflorencie_2016}, opening the way to the theory of tensor networks \cite{Verstraete01032008,Collura_2024}, and in understanding quantum quenches and the mechanisms leading to local relaxation in closed quantum systems \cite{calabrese2005evolution,alba1,alba2,Kaufman_2016}.

More recently, several other quantum-information-inspired quantities have attracted attention as probes of properties that are not captured by entanglement alone. These include, among others, quantum magic \cite{Bravyi2005,Veitch_2014,howard2017,Leone2022}, which characterizes the departure of a state from the stabilizer set and thereby quantifies the resources required to simulate it through stabilizer-based techniques; non-Gaussianity \cite{genoni2008,genoni2010,marian2013}, which characterizes the departure of a state from the manifold of Gaussian states and hence the breakdown of a description based on Wick's theorem; athermality, which quantifies the departure of a state from thermal equilibrium \cite{Brandao2013Resourcetheory,Lostaglio2015,gour2015}; coherence, which quantifies the degree to which a quantum state is not diagonal with respect to a certain basis in a Hilbert space \cite{Baumgratz2014coherence, Streltsov2017coherence, Saxena2020coherence}; and entanglement asymmetry, which probes the symmetry properties of a state beyond what can be inferred from expectation values alone \cite{brs-07, vaccaro2008tradeoff, Gour:2009abc,Marvian:2014awa,ares2023entanglement}. 

A common language for organizing these quantities is provided by quantum resource theories \cite{chitambar2019quantum}. A resource theory is specified by a set of free states and a set of free operations, namely transformations that preserve the free set and therefore cannot generate the resource from a free state. States outside the free set are regarded as resourceful. Entanglement provides the paradigmatic example, with separable states as free states and local operations and classical communication (LOCC) as free operations. The other quantities mentioned above can be placed within analogous frameworks: in the resource theory of magic, the free states are stabilizer states and the free operations are Clifford operations \cite{Veitch_2014}; for non-Gaussianity, they are Gaussian states and Gaussian operations \cite{genoni2008}; for athermality, they are Gibbs states and thermal operations \cite{Brandao2013Resourcetheory,Lostaglio2015}; for coherence, free states are diagonal density matrices and the free operations do not introduce off-diagonal elements; and for asymmetry, free states are symmetric states and free operations commute with the symmetry generators \cite{Gour:2009abc,Marvian:2014awa}.

An important role within a resource theory is played by resource monotones, namely functions that vanish on free states and cannot increase under free operations. This property makes it possible to compare states according to their resource content and constrains which transformations of a state are achievable with free operations only. At the same time, a quantity does not need to satisfy all the requirements of a resource monotone in order to provide a useful characterization of a state. In many-body physics, analytically accessible diagnostics often reveal important properties of a state or of an evolution protocol even when they are not ``good monotones''  in the sense specified above. It is therefore useful to distinguish between two complementary roles: detecting a property of a quantum state, and quantifying it in a manner consistent with the transformations allowed by a resource theory.
This distinction is already familiar from entanglement theory. For pure bipartite states, the entanglement entropy provides a natural and essentially complete characterization of entanglement under LOCC. For mixed states, the situation is richer and different measures, such as the entanglement negativity \cite{VidalPRA2002,Plenio2005}, capture different aspects of the entanglement structure. The coexistence of several quantifiers simply reflects different questions one may wish to ask.

The interplay between these viewpoints has recently become particularly relevant in the study of entanglement asymmetry. Given a density matrix $\rho$, pure or mixed, and an abelian charge $Q$, one introduces the symmetrized state \begin{equation} \rho_{Q}=\sum_q\Pi_q\rho\Pi_q, \end{equation} where $\Pi_q$ projects onto the eigenspace of $Q$ with eigenvalue $q$. The state $\rho_{Q}$ is obtained by removing the coherences between different charge sectors and is symmetric under the transformations generated by $Q$. The von Neumann entanglement asymmetry can then be expressed as \begin{equation} \Delta S = S(\rho_{Q})-S(\rho) = S(\rho\Vert\rho_{Q}). \label{eq:von_neumann_asymmetry} \end{equation}
If one is interested in the description of the asymmetry in a region $A$ of the physical system, where the reduced density matrix is $\rho_A =\Tr_{\Bar{A}}{\rho}$ and the charge operator is $Q_A$, then simply:
\begin{equation} \Delta S_A = S(\rho_{A,Q})-S(\rho_A) = S(\rho_A\Vert\rho_{A,Q})\,, \quad \rho_{A,Q}=\sum_q\Pi_q\rho_A\Pi_q. \label{eq:von_neumann_asymmetry_subsystem} \end{equation}
As we discuss in Sec. \ref{sec:monotones}, expressing the von Neumann entanglement asymmetry 
as a relative entropy in equations \eqref{eq:von_neumann_asymmetry} and \eqref{eq:von_neumann_asymmetry_subsystem} makes explicit its status as a monotone in the resource theory of asymmetry.  Entanglement asymmetry has provided a particularly fruitful tool for investigating the breaking and the restoration of symmetries in quantum many-body settings \cite{ ares2023entanglement,ares2023lack,rylands2024microscopic,bertini2024dynamics,rylands2024dynamical,klobas2024non,ferro2024non, klobas2024translation,chalas2026from}, in random states and random quantum circuits \cite{capizzi2024universal, Ares2024BlackHole, chen2024ph, Ares25,russotto2025, russotto25u1, Joshi2026, castroalvaredo2026automata}, quantum and conformal field theories \cite{Casini:2019kex,Casini:2020rgj,magan2021,capizzi2023asymmetryIsing, chen2024renyi, fossati2024entanglement, benini2025entanglementCFT, Kusuki2024, fossati2024, florio2025,fujimura2025entanglement,benedito2026}, in systems presenting generalized and higher-order symmetries \cite{benini2025entanglemenHigher, benini25cat, GattoLamas25} and, recently, in connection with the resource theory of non-Gaussianity \cite{Ares2026lowerbound}. The dynamics of $U(1)$-asymmetry has also been experimentally observed in out-of-equilibrium quantum simulations \cite{joshi2024observing,Xu2025Observation, Joshi2026}.

A prominent phenomenon in the dynamics of quantum systems undergoing a restoration of some initially broken symmetry is the quantum Mpemba effect. This effect occurs when states that initially break a symmetry more strongly can restore it faster than states with a smaller initial asymmetry \cite{ares2023entanglement,rylands2024microscopic, murciano2024entanglement, chalas2024multiple,  caceffo2024entangled,foligno2024non, shion2, yamashika2024quenching,turkeshi2024quantum, liu2024symmetry,liu2024mpembaloc, banerjee2024asymmetry, summer2025resource,yu2025symmetry,aditya2025mpemba,Di_Giulio_2025,aditya2026FragmentedMpemba2026,MpembaReview,Calabrese_2026}. The resource-theoretic interpretation of the von Neumann asymmetry gives a precise meaning to this comparison: the initial states are ordered according to a monotone of asymmetry, and the subsequent dynamics describes the relaxation of this resource. 

In many-body applications, the study of entanglement asymmetry has been substantially advanced by introducing a family of Rényi generalizations. For a real and positive Rényi index $\alpha$, the Rényi entanglement asymmetry is conventionally defined as \cite{ares2023entanglement,murciano2024entanglement} \begin{equation} \Delta S^{(\alpha)} = S_\alpha(\rho_{Q})-S_\alpha(\rho), \qquad S_\alpha(\rho) = \frac{1}{1-\alpha}\log\mathrm{Tr}\,\rho^\alpha . \label{eq:renyidiff} \end{equation} The subsystem Rényi asymmetry $\Delta S_A^{(\alpha)}$ is obtained  replacing $\rho$ and $\rho_Q$ with $\rho_A$ and $\rho_{A,Q}$ respectively. This definition has several important properties. It is non-negative, it vanishes if and only if $\rho_A$ is symmetric, and it reduces to the von Neumann entanglement asymmetry in the limit $\alpha\to1$. Moreover, for integer values of $\alpha$, it can be evaluated in terms of charged moments using replica techniques. These features have made $\Delta S_A^{(\alpha)}$ a powerful and versatile diagnostic tool, enabling analytic investigations of symmetry restoration and the quantum Mpemba effect in a broad range of many-body systems. From the resource-theoretic perspective, however, such quantities do not constitute good asymmetry monotones \cite{kusuki2026resource}, and so in principle they do not allow for the same level of interpretation which is carried by the relative entropy \eqref{eq:von_neumann_asymmetry}. 

Equation~\eqref{eq:von_neumann_asymmetry}, however, reveals that there are two natural routes to generalizing the von Neumann entanglement asymmetry. One may extend its representation as an entropy difference, leading to Eq.~\eqref{eq:renyidiff}, or one may seek a Rényi extension that preserves the monotonicity of its relative-entropy representation. The latter route leads to quantities that retain a direct resource-theoretic interpretation. We introduce their precise definitions and discuss their properties in Sec.~\ref{sec:monotones}.
The two routes preserve different useful features of the von Neumann quantity. The Rényi entropy difference, in light of its simple definition, is naturally suited to charged-moment and replica techniques. The monotone extensions preserve the ordering of states induced by the allowed operations of the resource theory, at the expense of a representation in terms of charged moments which is not as straightforward. Both reduce to the same von Neumann asymmetry as $\alpha\to1$, but they need not coincide at different Rényi indices. Indeed, the equality between the entropy difference and the relative entropy in Eq.~\eqref{eq:von_neumann_asymmetry} is a special property of the von Neumann case and does not generally persist under Rényi generalization. These observations raise a natural question: to what extent do the different Rényi extensions lead to the same physical picture of asymmetry and its dynamics? This question is especially relevant for the quantum Mpemba effect. If the Rényi entropy difference and the monotone extensions produce the same ordering and relaxation behaviour, the phenomenology identified through $\Delta S^{(\alpha)}$ also acquires an indirect resource-theoretic interpretation. If they differ, their comparison can reveal which aspects of the dynamics are robust across different quantifiers and which reflect the particular Rényi extension under consideration.

The purpose of this work is to investigate this relation systematically. We consider Rényi asymmetry quantifiers that are proper monotones under the relevant symmetric operations. Importantly, we show that these monotones can be determined from the same charged moments that enter the computation of $\Delta S^{(\alpha)}$. The resource-theoretic formulation can thus be combined with the analytic and computational techniques that have made the Rényi entropy difference particularly effective in many-body applications. We find that the two families of quantities exhibit the same qualitative behaviour in several physically relevant regimes. This agreement explains why the Rényi entropy difference successfully captures a broad range of phenomena associated with asymmetry and provides a resource-theoretic interpretation of the corresponding results. At the same time, we identify minimal settings in which the two constructions lead to significantly different behaviours. These examples clarify the relation between the two Rényi extensions and illustrate the additional information that follows from imposing monotonicity. Rather than selecting one quantifier over the other, our analysis delineates the information encoded by each of them and establishes a bridge between the operational framework of quantum resource theories and the charged-moment methods used in quantum many-body physics.

The paper is organized as follows. In Sec.~\ref{sec:monotones}, we introduce the Rényi asymmetry quantifiers considered throughout this work and discuss their main properties. In Sec.~\ref{sec:replicatrick}, we show how these quantities can be extracted from charged moments via a two-variable replica trick, and apply this framework to integrable quenches and Haar-random states, uncovering a remarkable connection with the Rényi difference. In Sec.~\ref{sec:universality}, we demonstrate how these results support the emergence of a universal Mpemba effect across the entire family of asymmetry probes in quantum quenches of homogeneously tilted ferromagnetic states. We then present a simple toy model, in Sec \ref{sec:swap_model}, which allows us to study the dynamics of all asymmetry quantifiers across a range of different initial states. In Sec.~\ref{sec:matters}, we show how the two classes of quantifiers can lead to qualitatively different results when the initial state is not homogeneous. Finally, we summarize our findings and present our conclusions in Sec.~\ref{sec:conclusions}.

\section{R\'enyi divergences as good asymmetry monotones} \label{sec:monotones} We consider a quantum state $\rho$, which may be either pure or mixed, together with a global $U(1)$ charge $Q$. Examples include spin-$1/2$ chains, where $Q$ corresponds to the total magnetization, and fermionic systems, where $Q$ is the total particle-number operator. In this setting, the von Neumann entanglement asymmetry quantifies the amount of $U(1)$-symmetry breaking carried by a state exactly as specified by Eqs. \eqref{eq:von_neumann_asymmetry} and \eqref{eq:von_neumann_asymmetry_subsystem}. As anticipated in Sec. \ref{sec:intro}, the quantity $\Delta S (\rho)=S(\rho||\rho_Q)$ is a good asymmetry monotone, meaning that it vanishes on all and only the symmetric states $(\rho=\rho_Q)$ and it cannot increase under any $U(1)$-symmetric operation applied to the state $\rho$. The latter property, in particular, is a consequence of the quantum data processing inequality (QDPI) satisfied by the relative entropy \cite{nielsen2010quantum}: for any completely positive and trace-preserving quantum channel $\Lambda$, $S(\Lambda(\rho)\Vert\Lambda(\sigma)) \le S(\rho \Vert \sigma)$. Indeed, if $\Lambda$ is a $U(1)$-covariant quantum channel, such that $\Lambda (U_\alpha \rho U_\alpha^\dagger)=U_\alpha \Lambda (\rho) U_\alpha^\dagger$, for $U_\alpha =e^{i\alpha Q}$, then $\Lambda(\rho_Q)=[\Lambda(\rho)]_Q$ and the QDPI immediately implies: 
\begin{equation}
    \Delta S(\Lambda(\rho)) \le \Delta S(\rho).\label{eq:monotonicity}
\end{equation} 
Here, we focus on the case of $U(1)$-entanglement asymmetry, but some of our considerations hold more generally.  

Although the relative entropy \eqref{eq:von_neumann_asymmetry} is a good monotone, it is a formidable task to directly compute it, even in simple models. For this reason, Refs. \cite{ares2023entanglement,murciano2024entanglement} introduced replicated versions of the relative entropy through \eqref{eq:renyidiff}, exploiting the property that the relative entropy of asymmetry separates in the difference of two von Neumann entropies. This replica generalisation is particularly fruitful, as the R\'enyi difference \eqref{eq:renyidiff} can be recast as an integral of charged moments \cite{Goldstein-sela,ares2023entanglement}, which can often be evaluated analytically \cite{murciano2024entanglement,rylands2024microscopic}.
At the same time, for $\alpha\neq1$, the difference of $\alpha$-Rényi entropies does not generally define an asymmetry monotone, and may increase under symmetric operations. For example, an explicit $\mathbb Z_2$-symmetric protocol increasing $\Delta S^{(2)}$ was constructed in Ref.~\cite{kusuki2026resource}, and related examples are known in other settings. Since the limit $\alpha\to1$ is often difficult to access directly, the quantities $\Delta S^{(\alpha)}$ are frequently employed as probes of symmetry breaking in their own right. This naturally raises the question of to what extent conclusions drawn from these quantities are reproduced by asymmetry measures that possess the monotonicity property of Eq.~\eqref{eq:monotonicity}.

In the context of asymmetry dynamics and symmetry restoration, it is natural to ask whether a quantum Mpemba effect identified through a given asymmetry quantifier reflects a genuine reversal in the asymmetry content of the evolving states. Let $\mathcal O$ be a quantity used to characterize asymmetry. Given two initial states $\rho(0)$ and $\rho'(0)$ undergoing the same symmetric evolution, quantum Mpemba effect is said to occur if
\begin{equation}
    \mathcal{O}(\rho(0))> \mathcal{O}(\rho'(0)),\hspace{0.5cm} \mathcal{O}(\rho(t))< \mathcal{O}(\rho'(t)),
\end{equation} 
for all sufficiently large times $t$. When $\mathcal O$ is an asymmetry monotone, such an inversion admits a direct interpretation in resource-theoretic terms, since the quantity provides a faithful measure of asymmetry along the evolution. This makes monotones particularly appealing probes for studying Mpemba phenomena. At the same time, other asymmetry indicators, including the Rényi differences, may remain highly informative and practically useful, especially given their analytical accessibility. Motivated by this perspective, it is worthwhile to develop Rényi asymmetry quantifiers that both reproduce Eq. \eqref{eq:von_neumann_asymmetry} in an appropriate replica limit and retain the monotonicity properties of the relative entropy of asymmetry.

When trying to generalize a relative entropy to a Rényi version, different possibilities arise owing to the non-commutativity of general density matrices. Several different options have been introduced, most notably the Petz Rényi divergence \cite{PETZ198657} and the ``sandwiched'' Rényi divergence \cite{muller2013,Casini:2018cxg}. In this work we focus on the latter, defined as
\begin{equation}
\label{eq:sandwiched_R\'enyi_def}
    S^{(\alpha)} (\rho||\sigma) = \frac{1}{\alpha-1} \log \Tr \left[ \left(\sigma^{\frac{1-\alpha}{2\alpha}} \rho \sigma^{\frac{1-\alpha}{2\alpha}}\right)^\alpha\right],\hspace{0.5cm}\alpha >0.
\end{equation}
It is not difficult to convince oneself that in the limit $\alpha \to 1$, this quantity reduces to the standard relative entropy $S(\rho||\sigma)$. Moreover, $S^{(\alpha)} (\rho||\sigma)$ satisfies the QDPI for $\alpha \in [1/2,\infty)$, and thus it provides an asymmetry monotone in the corresponding range of values of $\alpha$, making it an efficient generalization of the entanglement asymmetry in multi-replica situations. In particular, in the special case $\alpha=1/2$, $S^{(1/2)} (\rho||\sigma)=-2\log F(\rho,\sigma)$, with $F$ the quantum state fidelity. Finally, $S^{(\alpha)}(\rho||\sigma)$ is a monotonically increasing function of $\alpha$ \cite{muller2013}.

Due to the negative exponent of $\sigma$ in Eq. \eqref{eq:sandwiched_R\'enyi_def} when $\alpha >1$, a charged-moment representation of $S^{(\alpha)}(\rho || \rho_Q)$ is not straightforward to obtain; we will solve this problem in the next section through a two-parameter replica trick. At this stage, we note that it is simple to evaluate this quantity on a pure state $\rho$. Indeed, given $\rho=\ket{\psi}\bra{\psi}$, the ket can be decomposed in the eigenstates of the charge as $\ket{\psi}=\sum_q \sqrt{p_q}\ket{\psi_q}$, with the states $\ket{\psi_q}$ chosen to be orthonormal. Exploiting this decomposition, the difference of $\alpha$-Rényi entropies evaluates to
\begin{equation}
\Delta S^{(\alpha)} (\rho) = S_\alpha (\rho_Q) = H_\alpha (\{p_q\})\,,   
\end{equation}
where $H_\alpha(\{p_q\})= (1-\alpha)^{-1}\log \sum_q p_q^\alpha$ is the classical $\alpha$-Rényi entropy of the charge probability distribution $\{p_q\}$. On the other hand, the sandwiched divergence yields
\begin{equation}
\label{eq:purestatesAlphato1overAlpha}
S^{(\alpha)}(\rho || \rho_Q) = \frac{\alpha}{\alpha-1} \log \sum_q p_q^{1/\alpha} = H_{1/\alpha}(\{p_q\})\,,
\end{equation}
implying the following exact relation for pure states:
\begin{equation}
    S^{(\alpha)}(\rho || \rho_Q) = \Delta S^{(1/\alpha)}(\rho). \label{eq:relationpure}
\end{equation}
One of the main results of this work, which will be discussed in the following sections, is that this equality approximately holds also in relevant setups where the state is not pure, namely homogeneous quenches and Haar-random states. Therefore, in these settings $ \Delta S^{(\alpha)}(\rho)$ exhibit the same qualitative behaviour as a good monotone. At this stage, we note that the equality \eqref{eq:relationpure} implies that for pure states the difference of Rényi entropies $\Delta S^{(\alpha)} (\rho)$ 
inherits monotonicity for pure-to-pure symmetric transformations in the relevant range. However, a general covariant quantum channel may transform a pure state into a mixed state, where the identity no longer holds.

The mapping of both $\Delta S^{(\alpha)} (\rho)$ and $S^{(\alpha)}(\rho||\rho_Q)$ to classical Rényi entropies when $\rho$ is a pure state allows for a clear physical interpretation of how the two quantities account for charge fluctuations. For a pure state, when $\alpha >1$, $S^{(\alpha)}(\rho||\rho_Q)$ is an entropic measure sensitive to the \textit{tail} of the charge probability distribution, while $\Delta S^{(\alpha)} (\rho)$ is only sensitive to the \textit{bulk} probability; when $\alpha <1$, the situation is reversed. The fact that Rényi entropies $H_\alpha(\{p_q\})$ provide different information on the probability distribution when the index $\alpha$ is varied is known in the literature. In particular, Rényi entropies with different values of $\alpha$ are necessary to fully describe the equilibrium probability distribution of systems with a multifractal structure \cite{jizba2004world}, for which the Shannon entropy is not sufficient. In appendix \ref{app:bulk_tail}, we provide a simple example that illustrates the different behaviour of $H_\alpha(\{p_q\})$ according to the value of $\alpha$.

\section{A replica trick for Rényi divergences}
\label{sec:replicatrick}
In this section, we develop a replica approach for the Rényi divergences, which is particularly fruitful in the application to entanglement asymmetry. We consider the object
\begin{equation}
\label{eq:replica_trick}
    S^{(m,n)} (\rho||\rho_Q) = \log \Tr \left[ \left(\rho_Q^{m/2} \rho \rho_Q^{m/2}\right)^n\right],
\end{equation}
where $m,n$ are integers. The $\alpha$ Rényi divergence is then obtained through the analytic continuation 
\begin{equation}
\label{eq:physical_limit}
    m\to \frac{1-\alpha}{\alpha}, \hspace{0.2cm}n\to \alpha.
\end{equation} This is a two-variable replica trick, where the relevant ``physical'' limit involves the subspace $n(m+1)=1$. If $m, n$ are integers, the quantity in \eqref{eq:replica_trick} admits the considerable simplification
\begin{equation}
   \log \Tr \left[ \left(\rho_Q^{m/2} \rho \rho_Q^{m/2}\right)^n\right] =  \log \Tr \left[\prod_{j=1}^n  \rho \rho_Q^{m}\right],
\end{equation}
therefore removing all issues arising from the presence of fractional powers of the symmetrized density matrix. 
Importantly, such simplification allows to express \eqref{eq:replica_trick} in terms of charged moments, which are commonly used in the context of entanglement asymmetry to evaluate $\Delta S^{(\alpha)}$. In particular, expressing the symmetrized state through the integral representation of the projectors over the eigenspaces of a $U(1)$ charge \cite{ares2023entanglement},
\begin{equation}
    \rho_Q = \sum_q \Pi_q \rho \Pi_q= \int \frac{d\alpha}{2\pi} e^{i\alpha Q} \rho e^{-i\alpha Q},
\end{equation}
the products of density matrices can be expanded in full generality as
\begin{equation}
\begin{split}
     S^{(m,n)}(\rho||\rho_Q) &= \log \int \prod_{j=1}^n\prod_{k=1}^m \frac{d\lambda_{j,k}}{2\pi}\Tr\left[\prod_{j=1}^n \rho \prod_{k=1}^m e^{i\lambda_{j,k}Q}\, \rho \, e^{-i\lambda_{j,k} Q}  \right] \\&=  \log \int \prod_{j=1}^n\prod_{k=1}^m \frac{d\lambda_{j,k}}{2\pi} \mathcal{Z}_{n,(m+1)} (\vec{ \lambda}_{1}, \dots \vec{ \lambda}_{n}),
    \label{eq:chargedmoments}
\end{split}
\end{equation}   
where $\mathcal{Z}_{n,(m+1)} (\vec{ \lambda}_{1}, \dots \vec{ \lambda}_{n})$ is the charged moment of index $N=n(m+1)$ evaluated at the parameters $\vec{\lambda}_j = (\lambda_{j,1},\lambda_{j,2} - \lambda_{j,1},\dots ,\lambda_{j,m}-\lambda_{j,m-1},-\lambda_{j,m} ).$
Compared to the standard case of \cite{ares2023entanglement}, the charged moments have additional structure arising mainly from the product in $j$. 
For example, for $n=2$ and $m=1$ the charged moment reads
\begin{equation}
    \mathcal{Z}_{2,2}(\lambda_1, -\lambda_1, \lambda_2, - \lambda_2)= Z_{4}(\lambda_1, -\lambda_1, \lambda_2, - \lambda_2).
\end{equation}
where we have introduced the ``standard'' charged moment \cite{Goldstein-sela}
\begin{equation}
\label{eq:standard_charged_moments}
    Z_n(\beta_1,\dots \beta_n) = \Tr\left[\prod_{j=1}^n\rho e^{i \beta_j Q}  \right].
\end{equation}
The structure of the vector $\vec{\lambda}_j$ can be made more symmetric by introducing two additional parameters $\lambda_0 = \lambda_{m+1}=0$, leading to: 
\begin{equation}
   \vec{\lambda}_j = (\lambda_{j,1} - \lambda_{j,0},\lambda_{j,2} - \lambda_{j,1}\dots, \lambda_{j,m+1}-\lambda_{j,m} ).  \label{lambdalist}
\end{equation}
In this way, the charged moment is very close to that appearing in the evaluation of $\Delta S^{(\alpha)}$, but with the extra parameters $\lambda_{j,0}, \lambda_{j,m+1}$ that are not integrated over (being identically zero) and with $n$ decoupled families of parameters $\vec \lambda_j$.
This is remarkable, as it shows that the evaluation of Rényi divergences involves the same ingredients required for the evaluation of $\Delta S^{(\alpha)}$.
In a generic setting, solving the integrals of the charged moments and taking the physical limit for $m$ and $n$ is out of reach both analytically and numerically (this is also the case for the standard charged moments \eqref{eq:standard_charged_moments}). However, for the quenches which have been commonly considered in the literature, the quasiparticle arguments for free systems \cite{calabrese2020entanglement,rottoli2024entanglementhamiltoniansquasiparticlepicture} and space-time duality for interacting integrable systems \cite{bertini2022growth,bertini2025exactly,travaglino2026space, rylands2024microscopic} allow to make considerable progress. Once the integrals have been cast into a more manageable form, a saddle point solution can be derived in the hydrodynamic limit. Moreover, in Haar random states the evaluation of \cite{Ares2024BlackHole,russotto2025} can be directly generalised to the present scenario. Below, we discuss in detail two relevant cases in which we can compute the quantities \eqref{eq:chargedmoments}.

\subsection{Quenches from solvable initial states}
We focus on a quench in a model of one-dimensional spinless free fermions, described by the hopping Hamiltonian \begin{equation}
\label{eq:hopping_hamiltonian}
    H =-\frac{1}{2}\sum_j c_j^\dag c_{j+1}+h.c. = \sum_k\varepsilon_k \, c_k^\dag c_k ,
\end{equation}
where $\varepsilon_k = -\cos k$. The system is initialized in the squeezed state
\begin{equation}
\label{eq:initialsqueezed}
\ket{\psi}\propto \exp \left\{\int_{k>0} \frac{dk}{2\pi} M(k) c_k^\dag c_{-k}^\dag \right\} \ket{0},
\end{equation}
which generalises the tilted ferromagnetic family commonly  used in the literature on entanglement asymmetry; such states are frequently considered in free quenches as they arise when the pre- and post- quench Hamiltonians are related by Bogoliubov transformations.  The occupation functions of this state are related to the function $M(k)$ through $\vartheta(k) = \frac{|M(k)|^2}{1+|M(k)|^2}$. These are typically parametrised through the Bogoliubov angle $\Delta_k$ as
\begin{equation}
    \vartheta(k) = \frac{1}{2}(1+\cos \Delta_k).
\end{equation}
States in the tilted ferromagnetic family\footnote{Properly speaking, the tilted ferromagnetic state is defined in the spin representation as $\ket{\theta} = \exp\left\{\frac{i}{2} \theta \sum_j \sigma_j^x\right\} \ket{\uparrow\uparrow \dots}$. Since this state is not a standard Gaussian state in the fermionic representation, one considers the cat version $\frac{1}{\sqrt{2}}(\ket{\theta}+\ket{-\theta})$: this is almost Gaussian, in the sense that the non-Gaussian features vanish for large $L$. It is precisely the cat state in the large $L$ regime which is obtained by choosing the Bogoliubov angle as in \eqref{eq:occTilted}.}, characterized by a tilting angle $\theta$, are obtained as a special case of the state \eqref{eq:initialsqueezed} when $\Delta_k$ has the following expression in terms of $\theta$ \cite{murciano2024entanglement}:
\begin{equation}
\label{eq:occTilted}
   \cos \Delta^{(\theta)}_k = \frac{ 2\cos\theta-(1+\cos^{2}\theta)\cos k }{ 1-2\cos\theta\cos k+\cos^{2}\theta }.
\end{equation}

In the study of quantum quenches, one is interested in the time-dependent properties of the reduced density matrices (RDM) of a subsystem, namely $\rho_A(t)  = \Tr_{\overline{A}}[\rho(t)]$, with $\overline{A}$ the complement of subsystem $A$. We will focus on subsystems of size $\ell$, in a chain of total length $L \to \infty$. All discussions of the previous section can be repeated by simply replacing the subsystem RDM in the expressions; for example the charged moments are defined at the level of the reduced state as 
\begin{equation}
    Z^A_n(\lambda_1,\dots \lambda_n) = \Tr\left[\prod_{j=1}^n\rho_A(t) e^{i \lambda_j Q_A}  \right],
\end{equation}
where the subsystem charge $Q_A$ is given by the sum of local charged densities restricted to sites in $A$.
In quenches from squeezed states, the subsystem charged moments admit an expansion in the hydrodynamic limit $t,\ell \gg 1$ of the form  \cite{ares2023entanglement, murciano2024entanglement,Travaglino2025measurements} 
\begin{equation}
\label{eq:chargedmomentssolvable}
    Z^A_N (\boldsymbol{\lambda}) =   Z^A_N (0)  e^{ \int \frac{dk}{4\pi} \ell_k(t) \sum_{j}f(\lambda_j)},
\end{equation}
where we have introduced the counting function $\ell_k(t) = \max(\ell-2|\partial_k\varepsilon_k|t,0)$, representing the effective volume occupied by the quasiparticles that contribute to the charged moment, and the function 
\begin{equation}
    f(\lambda) =  \log (\vartheta(k) e^{2i \lambda} + (1-\vartheta(k))).
\end{equation}
Hereafter all momentum integrations are on the full Brilloin zone $[-\pi,\pi]$.
The factor $Z_N (0)$ represents the initial state contribution and can be factored out of the integral, contributing additively out of the logarithm. This term vanishes in the physical subspace $N=n(m+1) = 1$, since $Z_1(0) = \Tr[\rho_A]=1$. Therefore, the Rényi divergences are determined only by the exponential term in Eq. \eqref{eq:chargedmomentssolvable}. 

\begin{figure}[t!]
    \centering
    \includegraphics[width=.8\linewidth]{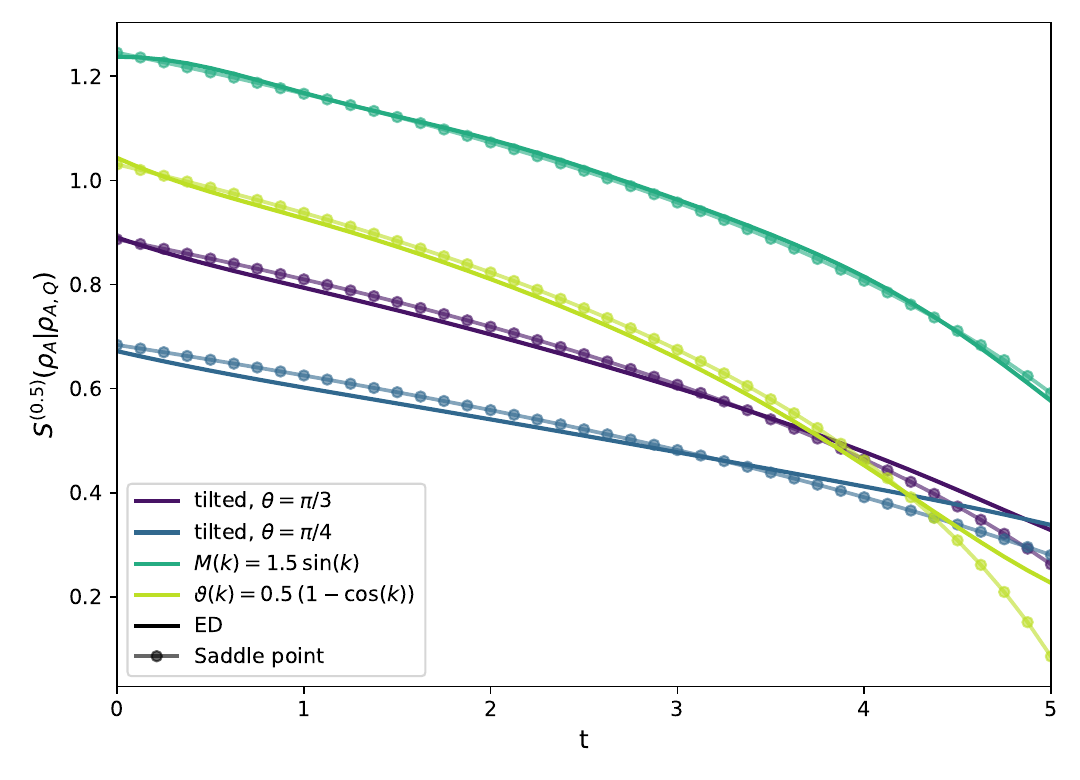}
    \caption{Comparison between exact diagonalization and the saddle point prediction \eqref{eq:asymptotic_divergence_smallt} for the sandwiched asymmetry $S^{(\alpha)}(\rho_A\Vert \rho_{A,Q})$ with $\alpha=0.5$  for several initial squeezed states constructed in different manners: the blue and the purple lines correspond to tilted ferromagnets with two choices of tilting angle. The two green lines are built by specifying the function $M(k)$ appearing in \eqref{eq:initialsqueezed} or directly the occupation function $\vartheta(k)$. The plot shows remarkably good agreement with the saddle point prediction even at very small system sizes ($L=20, \ell=10$); the prediction then starts to fail for increasing $t$, since $\ell_k(t)$ ceases to be large. }
    \label{fig:alpha05}
\end{figure}

\begin{figure}[t]
    \centering
    \includegraphics[width=.8\linewidth]{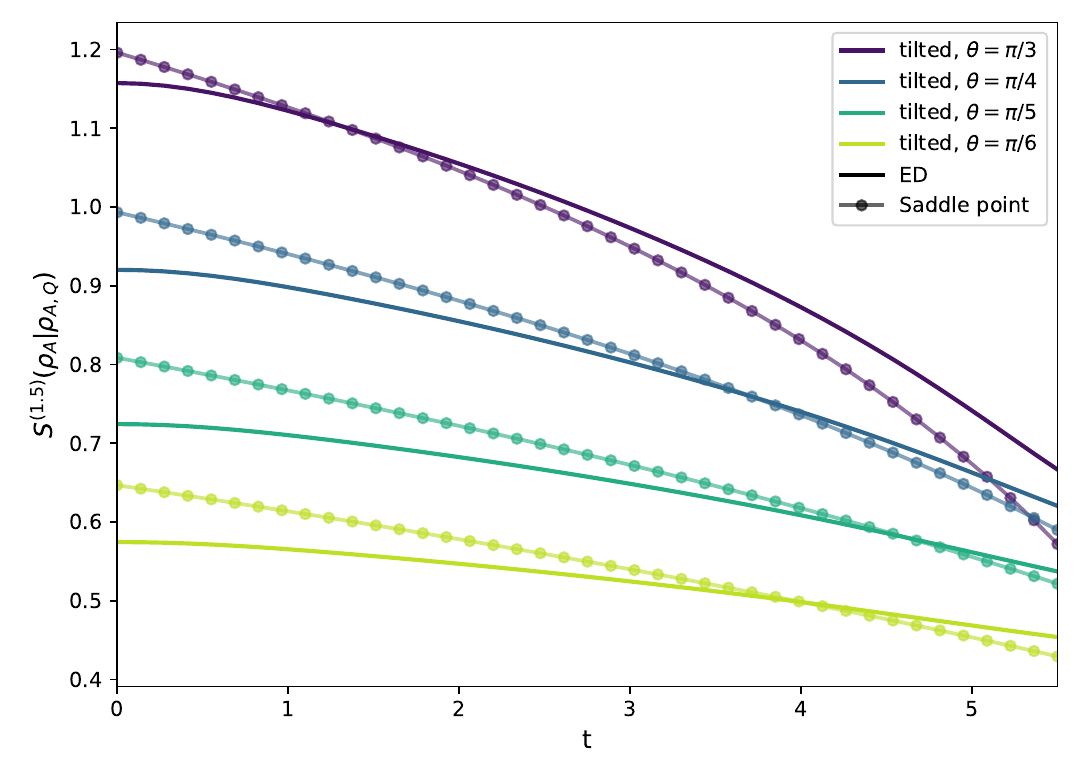}
    \caption{Comparison between exact diagonalization and the saddle point prediction \eqref{eq:asymptotic_divergence_smallt} for the sandwiched asymmetry $S^{(\alpha)}(\rho_A\Vert \rho_{A,Q})$ with $\alpha=1.5$, for different initial tilted ferromagnetic states at $\ell=10$. Although the comparison to the asymptotic result \eqref{eq:asymptotic_divergence_smallt} is not as good as the one in figure \ref{fig:alpha05}, the deviations are still controlled and compatible with the corrections $O(\ell_k^{-1})$ expected from the saddle point approximation.}
    \label{fig:alpha15}
\end{figure}

Given the structure \eqref{eq:chargedmomentssolvable} and the specific form \eqref{lambdalist} of the parameters, it is immediately clear that the dependence on the product over $j$ in Eq. \eqref{eq:chargedmoments} completely factorizes to a global power $n$. Hence, isolating the term $\log Z_{n(m+1)}(0)$, which vanishes in the replica limit, we obtain
\begin{equation}
\label{eq:Smn_freequench}
    S^{(m,n)}(\rho_A||\rho_{A,Q}) = n \log \int \prod_{k=1}^m \frac{d\lambda_k}{2\pi}  \exp\left\{\sum_{k=0}^{m}\int \frac{dk}{4\pi}\ell_k(t) f(\lambda_{k+1,k}) \right\} +  \log Z_{n(m+1)}(0),
\end{equation}
where we denoted $\lambda_{i,j} = \lambda_i - \lambda_j$. This expression does not allow to take the physical limit \eqref{eq:physical_limit} yet: we first need to solve the integrals in $\lambda_k$ by saddle point approximation. This is justified as long as we are interested in the hydrodynamic limit, in which both $\ell$ and $t$ are large, and in particular in the regime $t \ll \ell$, such that $\ell_k(t)\gg 1$ acts as a large parameter. It is easy to see that the saddle points are identical for all integration variables, $\lambda_k^* = 0\,\, \forall k$ . Taking the second derivative of $f(\lambda)$ at $\lambda=0$ yields $-4\,\vartheta(k) (1-\vartheta(k))$, and the saddle point evaluation consequently gives, neglecting the irrelevant factor $\log Z_{n(m+1)}(0)$:
\begin{equation}
     S^{(m,n)}(\rho_A||\rho_{A,Q}) \simeq -\frac{mn}{2} \log\left(\pi \int \frac{dk}{2\pi} \ell_k(t) \,\vartheta(k) (1-\vartheta(k)) \right) - \frac{n}{2} \log (m+1). 
\end{equation}
At this point the physical limit can be taken straightforwardly, yielding
\begin{equation}
\begin{split}
      S^{(\alpha)} (\rho_A||\rho_{A,Q}) &= \frac{1}{\alpha-1} \lim_{n\to \alpha} \lim_{m\to \frac{1-\alpha}{\alpha}} S^{(m,n)}(\rho||\rho_Q) \\
      &\simeq \frac{1}{2}\log\left(\pi \alpha^{\frac{\alpha}{\alpha-1}}\int \frac{dk}{2\pi} \ell_k(t) \,\vartheta(k) (1-\vartheta(k))\right). \label{eq:asymptotic_divergence_smallt}
\end{split}
\end{equation}
The comparison between the exact value of the Rényi divergence, obtained through exact diagonalization (ED) of the density matrix, and the asymptotic analytical prediction \eqref{eq:asymptotic_divergence_smallt} is shown in Fig. \ref{fig:alpha05} and Fig. \ref{fig:alpha15} for two values of $\alpha$. In both cases we observe excellent agreement, up to $O(\ell_k^{-1})$ corrections which are expected from the saddle point evaluation. 

Eq. \eqref{eq:asymptotic_divergence_smallt} is to be compared with the asymptotic behaviour of $\Delta S^{(\alpha)}$, given by \cite{murciano2024entanglement}
\begin{equation}
\label{eq:asymptoticR\'enyidiff}
    \Delta S^{(\alpha)}(\rho_A)  \simeq \frac{1}{2}\log\left(\pi \alpha^{\frac{1}{\alpha-1}}\int \frac{dk}{2\pi} \ell_k(t) \,\vartheta(k) (1-\vartheta(k))\right).
\end{equation}
Remarkably, the two expressions are related by the simple transformation $\alpha \leftrightarrow 1/\alpha$, which was discussed in section \ref{sec:monotones} as a general feature of pure states. The validity of such relation also in the present context, which involves mixed density matrices, can be understood heuristically by a simple hydrodynamic argument. Given a quench from a lowly entangled pure state, at large scales the state will be comprised of two parts~\cite{rottoli2024entanglementhamiltoniansquasiparticlepicture,travaglino2024}: on the edges of the subsystem, the state is mixed, but it is also symmetrised by the partial trace, and thus does not contribute to the total asymmetry of the state. In the bulk, the state is not symmetric, but by Lieb-Robinson bounds, it is pure up to exponentially small corrections. Therefore, the term which is actually contributing to the R\'enyi divergences is essentially pure, and it is reasonable to expect that the structure of the result retains (at leading order) the pure-state property \eqref{eq:purestatesAlphato1overAlpha}, as the explicit saddle point solution confirms. This argument is expected to hold whenever the initial state is pure and lowly entangled, and when the trace over the complement fully symmetrizes the state during the evolution. The latter condition is satisfied by quench dynamics which restores the symmetry at long times, and the only known counterexamples to this feature arise because of the presence of non-Abelian structures \cite{ares2023lack}. In Sec. \ref{subsec:latetimes}, we show that for such quenches the Rényi divergences can indeed be expressed as classical Rényi entropies of the charge fluctuations of the pure component of $\rho_A$ (at the hydrodynamic scale), thus confirming this heuristic insight.

We therefore conclude that in quenches from squeezed states the two Rényi asymmetry candidates are related by
\begin{equation}
    S^{(\alpha)}(\rho_A||\rho_{A,Q}) \simeq  \Delta S^{(1/\alpha)}(\rho_A),
    \label{eq:alphatooneoveralpha}
\end{equation}
where the $\simeq$ stands for equality at the hydrodynamic scale in the regime where the saddle point approximation is justified. Equation \eqref{eq:alphatooneoveralpha} is the main result of this section. This relation shows that the dynamics of the Rényi divergences and of the standard Rényi asymmetries is qualitatively (and quantitatively, at this scale) the same, as long as one is interested in hydrodynamic features. Thus, the differences of Rényi entropies, despite not being proper asymmetry monotones, behave as such for this common class of initial states.  Note that for $\alpha=0.5$, Eq. \eqref{eq:alphatooneoveralpha} states the asymptotic equivalence of the logarithm of the fidelity  and the difference of the Rényi-2 entropies, namely
\begin{equation}
2\log F(\rho_A,\rho_{A,Q}) \simeq S_2(\rho_A) - S_2(\rho_{A,Q})   .
\end{equation}

It is also interesting to consider the long-time behaviour of \eqref{eq:Smn_freequench}. In this regime, the most natural approach is to perform a Taylor expansion of the exponential, considering that $\ell_k(t)$ approaches zero for $t\to \infty$ as a consequence of symmetry restoration. Performing the expansion and integrating over the parameters $\lambda_k$, one obtains  
\begin{equation}
    S^{(m,n)}  \overset{t \to \infty}{\simeq} n(m+1)\int \frac{dk}{4\pi}\ell_k(t) \log g(k),
\end{equation}
where we have introduced $g(k) = \max(\vartheta(k),1-\vartheta(k))$. Since in the physical limit $n(m+1)=1$, it is clear that $S^{(\alpha)}(\rho_A\Vert\rho_{A,Q})$ depends on $\alpha$ only through the prefactor $(\alpha-1)^{-1}$, namely: 
\begin{equation}
\label{eq:divergencetlarge}
    S^{(\alpha)}(\rho_A||\rho_{A,Q}) \overset{t \gg \ell_A}{\simeq} \frac{1}{\alpha-1}  \int \frac{dk}{4\pi} \ell_k(t) \log g(k).
\end{equation}
Again, this is equivalent to the result for $\Delta S^{(1/\alpha)}$, since $\Delta S^{(\alpha)}$ has the following expression (see \cite{murciano2024entanglement}):
\begin{equation}
   \Delta S^{(\alpha)}(\rho_A) \overset{t \gg \ell_A}{\simeq} \frac{\alpha}{1-\alpha}  \int \frac{dk}{4\pi} \ell_k(t) \log g(k),
   \label{eq:differencelarget}
\end{equation}
confirming that the observed correspondence is not merely a feature of the short-time regime. In particular, it is straightforward to obtain that the decay to zero for large $t$ for both quantities follows the power law $t^{-3}$. Note however that, as it is, equation \eqref{eq:divergencetlarge} is problematic, as it becomes negative for $\alpha > 1,$ owing to the negativity of $  \log g(k)$, thus implying that it can be meaningful only in the opposite regime $\alpha<1$ (while for the same reason \eqref{eq:differencelarget} can be meaningful only in the regime $\alpha>1$, where it is usually considered).
This is not a completely unexpected feature, as the interplay between analytic continuation and limits to infinity may present some issues. In Sec. \ref{subsec:latetimes}, we provide a general construction, inspired by the approach of~\cite{rylands2024microscopic}, which allows us to find the limit $t\to \infty$ for all values of $\alpha$, and which confirms that the relation for $\alpha \to 1/\alpha$ remains valid in all scenarios. Moreover, while for $\alpha <1$ at large time $S^{(\alpha)}(\rho_A||\rho_{A,Q})$ decays as $t^{-3}$, for $\alpha>1$ we uncover a novel asymptotic time dependence, namely
\begin{equation}
    S^{(\alpha>1)}(\rho_A||\rho_{A,Q}) \overset{t \gg \ell_A}{\simeq} t^{-3/\alpha},
    \label{eq:dampeddecay}
\end{equation}
showing a freezing of the dynamics as $\alpha$ is increased. This is not surprising, because the $\alpha$-Rényi divergence for $\alpha=\infty$ simply counts the number of occupied charge sectors, and this decays logarithmically during the evolution at the ballistic scale. This result also implies that the difference of Rényi entropies satisfies $ \Delta S^{(\alpha<1)} \overset{t \gg \ell_A}{\simeq} t^{-3\alpha}$, a feature not previously noted in the literature.

Although we have focused on free systems, the main result of this section, Eq. \eqref{eq:alphatooneoveralpha}, generalises to interacting integrable models, for which the regime $t \ll \ell$ can be explored through space-time duality. 
In fact, it was proven in \cite{bertini2024dynamics} that the expression \eqref{eq:asymptoticR\'enyidiff} can be extended to integrable quenches, by substituting free quantities with the corresponding interacting analogues. The Rényi divergences admit a similar analysis, which leads to a confirmation of the $\alpha \leftrightarrow 1/\alpha$ relation. The details of this derivation are presented in appendix \ref{app:interacting}.

As a final comment, we note that since the hydrodynamic argument discussed previously does not rely, per se, on either limit $t \gg \ell$ or $t \ll \ell$, we expect the relation between Rényi divergences and Rényi asymmetries to hold also in the intermediate regime $t\sim\ell$, where neither the saddle-point approximation nor the large-$t$ expansion applies. We therefore conjecture, on the basis of the hydrodynamic argument, that \eqref{eq:alphatooneoveralpha} holds throughout the scaling regime $t,\ell\to\infty$ with $t/\ell$ fixed, for the evolutions considered here, in interacting integrable evolutions. This is explicitly shown to hold for free fermionic systems in Sec. \ref{sec:universality}.
 
\subsection{Haar random states}
A further situation in which the replica trick can be efficiently employed is the computation of the average Rényi divergences of a subsystem in a random state of qubits \cite{page1993average, page1993information,bianchi_22}. Remarkably, although random states are maximally chaotic, and have thus very different features from free fermionic states, we find that the simple mapping \eqref{eq:alphatooneoveralpha} works also in this context.

Let us consider a system of $L$ qubits, with computational basis $\{\ket{0}_k,\,\ket{1}_k\}$ on every site, and a subsystem $A$ consisting of $\ell$ sites. Our aim is to compute the average Rényi divergence $\mathbb{E}[S^{(\alpha)}(\rho_A||\rho_{A,Q})]$, where $\rho=\ket{\psi}\bra{\psi}$ and $\ket{\psi}=U\ket{0}$, $\ket{0}=\otimes_{k=1}^L\ket{0}_k$. The $2^L$ by $2^L$ matrix $U$ is randomly drawn from the group of unitary matrices, and averages over the group are computed using the Haar measure \cite{mele2024introduction}. Since the Haar random ensemble is self-averaging, the error committed in bringing the expectation value inside the logarithm is exponentially small in system size \cite{Ares2024BlackHole}. Thus, it is sufficient to compute
\begin{equation}
\mathbb{E}[M^{(m,n)}]=\mathbb{E}\left[\Tr\prod_{j=1}^n\rho_A \rho_{A,Q}^m\right]  = \int \prod_{j=1}^n\prod_{k=1}^m \frac{d\lambda_{j,k}}{2\pi} \mathbb{E}[\mathcal{Z}^{(A)}_{n,m+1} (\vec{ \lambda}_{1}, \dots \vec{ \lambda}_{n})]\,,
\end{equation}
and then take the logarithm of the above, before performing the physical limit. 

To do so, we need to slightly generalize the results of \cite{Ares2024BlackHole} on expectation values of standard charge moments. Let us consider the simplest case of $Z_2^{(A)}(\varphi_1,\varphi_2)$, for $\varphi_1,\,\varphi_2$ \textit{arbitrary}  phases in $[-\pi,\pi)$. Its expectation value is
\begin{equation}
\mathbb{E}[Z_2^{(A)}(\varphi_1,\varphi_2)]:= \mathbb{E}\left[\Tr\left(\rho_A e^{i\varphi_1 Q_A}\rho_A e^{i\varphi_2 Q_A}\right)\right]   = \bra{I^-(\varphi_1,\varphi_2)}_A\bra{I^+}_B\mathbb{E}[U^{\otimes 2}\otimes (U^*)^{\otimes 2}]  \ket{0}^{\otimes 4}\,,
\end{equation}
where we employed the Choi-Jamiolkowski representation (from a two- to a four-replica space) to map \cite{watrous2018theory} 
\begin{equation}
\rho \otimes \rho \mapsto \ket{\rho \otimes \rho}\equiv 1\otimes \rho\otimes1\otimes\rho\ket{I^+},
\end{equation}
  where $\ket{I^+}=\otimes_{k=1}^N\ket{I^+}_k$, $\ket{I^+}_k =\sum_{a_1,a_2=0}^1\ket{a_1a_1a_2a_2}_k$. Moreover, we introduced the states $\bra{I^+}_B = \otimes_{k\in B}\bra{I^+}_k$ and $\bra{I^-(\varphi_1,\varphi_2)}_A=\otimes_{k\in A}\bra{I^-(\varphi_1,\varphi_2)}_k$, with 
  \begin{equation}
  \bra{I^-(\varphi_1,\varphi_2)}_k=\sum_{a_1,a_2=0}^1e^{i(\varphi_1 a_2+\varphi_2 a_1)}\bra{a_1a_2a_2a_1}_k. 
  \end{equation} 
  This immediately translates to the $N$-replica case. For $\vec{\varphi}=(\varphi_1,\dots,\varphi_N)$, and defining (with $j+N\equiv j$ for the replica indices)
  \begin{equation}
  \bra{I^-(\vec{\varphi})}_k=\sum_{\{a_j=0\}}^1\otimes_{j=1}^N e^{i \varphi_j a_{j+1}}\bra{a_ja_{j+1}}_k,
  \end{equation} 
  one obtains:
\begin{equation}
\mathbb{E}[Z_N^{(A)}(\vec{\varphi})]=\bra{I^-(\vec{\varphi})}_A\bra{I^+}_B\mathbb{E}[U^{\otimes N}\otimes (U^*)^{\otimes N}]  \ket{0}^{\otimes 2N}\,.
\end{equation}
At this point, following a computation in the same spirit as that reported in \cite{Ares2024BlackHole} (using Weingarten calculus \cite{w-78,bb-96}), we obtain, up to terms exponentially suppressed in the system size,
\begin{align}
\mathbb{E}[Z_N^{(A)}(\vec{\varphi})] &\simeq \frac{1}{2^{NL}}\sum_{\sigma \in S_N} \left(\sum_{\{a_j=0\}}^1\prod_{j=1}^Ne^{i\varphi_j a_{j+1}}\delta_{a_{j+1},a_{\sigma(j)}}\right)^{\ell}\left(\sum_{\{a_j=0\}}^1\prod_{j=1}^N\delta_{a_{j},a_{\sigma(j)}}\right)^{L-\ell} \nonumber \\
&\simeq \frac{1}{2^{NL}}\left\{2^{(L-\ell)N}(1+e^{i\sum_j \varphi_j})^\ell + 2^{L-\ell + N\ell}\left(\prod_j e^{i\varphi_j/2}\right)^\ell \left[\prod_j \cos(\varphi_j/2)\right]^\ell\right\}\,.
\end{align}
In the first line above, $S_N$ is the set of permutations of $N$ indices, and in the second line we used the fact that, in the large-$L$ limit, the dominant permutations are $\sigma(j)=j\,\, \forall j$ and $\sigma(j)=j+1\,\, \forall j$. Moreover, we used  
\begin{equation}
\left(\sum_{\{a_j\}}\prod_je^{i\varphi_j a_{j+1}}\right)^\ell=\left[\prod_j(1+ e^{i\varphi_j})\right]^\ell=2^{N\ell}\left[\prod_j e^{i\varphi_j/2}\right]^\ell\left[\prod_j\cos(\varphi_j/2)\right]^\ell.    
\end{equation}

At this point, we can set $N=n(m+1)$ and specify the phases. These are $\varphi_1=\lambda_{1,1},\,\varphi_2=\lambda_{1,2}-\lambda_{1,1},\dots ,\varphi_{m+1}=-\lambda_{1,m},\varphi_{m+2}=\lambda_{2,1},\varphi_{m+3}=\lambda_{2,2}-\lambda_{2,1}\dots,\varphi_N = -\lambda_{n,m}$. In particular, they sum to zero, and thus:
\begin{equation}
\mathbb{E}[M^{(m,n)}] \simeq 2^{\ell[1-n(m+1)]}   + 2^{(L-\ell)[1-n(m+1)]} \int \prod_{j=1}^n\prod_{k=1}^m \frac{d\lambda_{j,k}}{2\pi} \left[\prod_{j=1}^{n(m+1)} \cos(\varphi_j/2)\right]^\ell\,.
\end{equation}
The above integral can be factorized and, after renaming the integration variables, it becomes
\begin{equation}
\left(\int\prod_{k=1}^m\frac{d\lambda_{k}}{2\pi} \exp \left\{\ell \sum_{k=1}^{m+1} \log \cos \left(\frac{\lambda_k -\lambda_{k-1}}{2}\right)\right\}\right)^n\,,
\end{equation}
where we set $\lambda_0= \lambda_{m+1}=0$. These variables select a unique maximum in the above exponent at $\lambda_k=0$ for all $k$. Therefore, after performing a simple saddle-point integration, we obtain
\begin{equation}
\mathbb{E}[M^{(m,n)}] \simeq 2^{\ell[1-n(m+1)]} +2^{(L-\ell)[1-n(m+1)]} \left(\frac{\pi \ell}{2}\right)^{-\frac{mn}{2}}(m+1)^{-\frac{n}{2}}\,.
\end{equation}
Finally, performing the physical limit as in Eq. \eqref{eq:asymptotic_divergence_smallt}, we obtain, at leading order in $\ell, L$
\begin{equation}
\label{eq:Haar_random_final}
\mathbb{E}[S^{(\alpha)} (\rho_A || \rho_{A,Q})] \simeq \frac{1}{\alpha-1} \lim_{n\to \alpha} \lim_{m\to \frac{1-\alpha}{\alpha}} \log \mathbb{E}[M^{(m,n)}] \simeq \begin{cases}
0, \quad &\ell <L/2 \\
\frac{1}{2}\log\left(\frac{\pi \ell}{2}\right) + \frac{\alpha \log \alpha}{2(\alpha-1)}, \quad &\ell >L/2
\end{cases}\,.
\end{equation}
Comparing with the result of \cite{Ares2024BlackHole}, we conclude that at leading order the mapping Eq. \eqref{eq:alphatooneoveralpha} holds also in the case of averaged subsystems Rényi divergences of Haar random states. This perhaps surprising result can be motivated physically by noting that, for $\ell < L/2$, the state $\rho_A$ is, on average, exponentially close to the infinite-temperature state (for which any reasonable measure of asymmetry must evaluate to zero). When $\ell > L/2$, however, we need a finer argument to explain the result Eq. \eqref{eq:Haar_random_final}, since it is not possible to argue that the state $\rho_A$ is pure (on average). As we show below, the result follows from the fact that $\rho_A$ admits an approximate factorization into a pure state carrying all the asymmetry, and a mixed state which does not contribute to $S^{(\alpha)}(\rho_A \Vert \rho_{A,Q})$. 

A pure random state in a bipartite Hilbert space $\mathcal{H}_A \otimes \mathcal{H}_B$ has the following general Schmidt decomposition \cite{tang2025blackhole,zyczkowski2001induced}
\begin{equation}
\ket{\psi}= \sum_{a=1}^{d_B} \sqrt{w_a} U\ket{e_a} \otimes \tilde{U}\ket{\tilde{e}_a},
\end{equation}
where $\ket{e_a}$ and $\ket{\tilde{e}_a}$ are orthonormal vectors in $\mathcal{H}_A$ and $\mathcal{H}_B$ respectively, and $U$, $\tilde{U}$ are random unitary matrices acting on the two spaces. The sum runs from $1$ to $d_B=2^{L-\ell}$ because, since $\ell > L/2$, the Schmidt rank is at most $d_B$. Let $\ket{u_a}= U\ket{e_a}$ and $\ket{v_a}=\tilde{U}\ket{\tilde{e}_a}$: these vectors are also obviously orthonormal, so the reduced density matrix reads
\begin{equation}
\rho_A = \sum_{a=1}^{d_B} w_a \ket{u_a}\bra{u_a}.
\end{equation}
We now introduce the standard projectors $\Pi_q$ onto the charge eigenspaces of $Q_A$, and the vectors 
\begin{equation}
\ket{e_{q,a}} = \frac{\Pi_q \ket{u_a}}{\sqrt{p_q}}, \quad p_q = \frac{\Tr \Pi_q}{d_A}=2^{-\ell} \binom{\ell}{q}.
\end{equation}
With these definitions, we can expand $\rho_A$ and $\rho_{A,Q}$ as
\begin{equation}
\label{eq:RDM_schmidt_expansion_complete}
\rho_A = \sum_{a=1}^{d_B} w_a\sum_{q,q'} \sqrt{p_q \,p_{q'}} \ket{e_{q,a}}\bra{e_{q',a}}, \quad \rho_{A,Q} = \sum_{a=1}^{d_B} w_a\sum_{q} p_q \ket{e_{q,a}}\bra{e_{q,a}}.
\end{equation}
The states $\ket{e_{q,a}}$ are not orthonormal, in general
\begin{equation}
\langle e_{q,a}|e_{q',a'}\rangle = \delta_{q,q'} (G_q)_{a,a'}, \quad (G_q)_{a,a'} = \frac{\bra{ u_a}\Pi_q \ket{u_{a'}}}{p_q}.
\end{equation}
This is where the fact that the matrix $U$ is a random unitary comes into play. Indeed, it is easy to compute the moments of the matrix $G_q$ \cite{mele2024introduction}. One finds that on average $\mathbb{E}[(G_q)_{a,a'}]=\delta_{a,a'}$, and the variances can be computed as well: $\mathbb{E}[|(G_q)_{a,a'}-\delta_{a,a'}|^2]= O(d_q^{-1})$ for all elements $a,a'$, where $d_q =\binom{\ell}{q}$ is the dimension of the charge eigenspace. Thus, for typical charges $q \simeq \ell/2$, the matrix $G_q$ is the identity up to exponentially small corrections (notice that it is also possible to find a uniform rather than entry-wise bound on the distance between $G_q$ and the identity). Because the probability distribution $\{p_q\}$ is binomial, the error committed in considering only eigenvalues in a small window around $q=\ell/2$ in the expressions \eqref{eq:RDM_schmidt_expansion_complete} is also exponentially small when $\ell$ is large. Therefore, the matrix elements of $\rho_A$ and $\rho_{A,Q}$ are given, for a fixed values of the charge and of the Schmidt index, by
\begin{equation}
(\rho_A)_{(q,a),(q',a')} \simeq w_a \sqrt{p_q p_q'}\delta_{a,a'}, \quad (\rho_{A,Q})_{(q,a),(q',a')} \simeq w_a p_q \delta_{a,a'} \delta_{q,q'},
\end{equation}
which means that these matrices admit the approximate (in the sense specified above) factorization
\begin{equation}
\label{eq:Haar_factorization}
\rho_A \simeq \ket{\phi}\bra{\phi}\otimes \tau, \quad \rho_{A,Q} \simeq \text{diag}(p_q)\otimes \tau,
\end{equation}
with
\begin{equation}
\ket{\phi}= \sum_q \sqrt{p_q}\ket{q}, \quad \tau =\sum_{a=1}^{d_B} w_a \ket{e_a}\bra{e_a},
\end{equation}
with $\ket{q}$ representative charge eigenstates. We remark that this is an effective factorization in a basis labeled by the two indices, not in a physical tensor-product Hilbert space. With these caveats, thanks to the fact that the sandwiched divergence is additive on tensor products, we obtain that for any typical random unitary realization
\begin{equation}
S^{(\alpha)}(\rho_A \Vert \rho_{A,Q}) \simeq S^{(\alpha)}\left(\ket{\phi}\bra{\phi}\bigg | \bigg | \sum_q p_q \ket{q}\bra{q}\right)=H_{1/\alpha}(\{p_q\}),
\end{equation}
and the expression \eqref{eq:Haar_random_final} follows from the Gaussian limit of the binomial probability distribution in the large-$\ell$ limit. The typical factorization Eq. \eqref{eq:Haar_factorization} explains also the average value of the standard difference of Rényi entropies $\Delta S^{(\alpha)}(\rho_A)$ found in \cite{Ares2024BlackHole} for $\ell \gg L/2$.

\section{Robustness of the Mpemba effect in spin chains}

\label{sec:universality}
In this section, we show that the results derived above imply the occurrence of the quantum Mpemba effect throughout the entire family of Rényi relative entropies, for all values of $\alpha$, in quenches from tilted ferromagnetic states, at least for integrable models.
The quantum Mpemba effect occurs when states that initially break a symmetry more strongly restore it faster than states with a smaller initial asymmetry. It is therefore characterized by a crossing of the entanglement asymmetry curves corresponding to quenches from two different initial tilting angles, followed by a reversal of their ordering at long times (with the possible exception of multiple crossings \cite{chalas2024multiple}).
The occurrence of crossings for the entire family of asymmetry monotones considered here, encompassing quantities related to both fidelity and relative entropy, represents a significantly stronger result than the observation of a crossing for a single measure of symmetry breaking, such as the entanglement asymmetry. Indeed, varying $\alpha$ changes the sensitivity of the monotones to different scales of the underlying fluctuations (see \cite{jizba2004world} and appendix \ref{app:bulk_tail}). A crossing that persists for all $\alpha$ can therefore be interpreted as being independent of the particular fluctuation scale probed by the quantifier. 

Generically the Mpemba effect occurs when given two initial states with different tilting angle $\theta_1$ and $\theta_2$, with initial asymmetries
\begin{equation}
    S^{(\alpha)}_{t=0}(\theta_2) > S^{(\alpha)}_{t=0}(\theta_1) 
\end{equation}
then for all $t$ larger than a given $t_M$, the order is reversed
\begin{equation}
    S^{(\alpha)}_{t>t_M}(\theta_2) < S^{(\alpha)}_{t>t_M}(\theta_1). 
\end{equation}
We say that the effect is robust when this happens for all $\alpha$. This property is by no means guaranteed. Indeed, numerous examples in the literature show that the occurrence of the Mpemba effect can depend strongly on the specific measure employed, with crossings appearing for some probes but being absent for others; see, e.g., Refs. \cite{Ares_2025_simpler,Ares_2025m,Russotto_2026,oac-26}.

\subsection{The initial R\'enyi divergences} 

To study the robustness of the Mpemba effect in the tilted ferromagnetic quench, we should analyze the two conditions above for short and long times.  

We begin by studying the ordering induced by the Rényi divergences at time zero. 
We consider the tilted ferromagnet for simplicity.
Indeed, although the Gaussian cat state used in the fermionic formulation for the dynamics has a mixed reduced density matrix, its entanglement asymmetry agrees with that of the tilted product state at leading order for large subsystems, with differences confined to subleading contributions. 
In this regime, since the state is pure, we have exactly
\begin{equation}
S^{(\alpha)}(\rho_A || \rho_{A,Q}) =  H_{1/\alpha}(\{p_q\})\,,
\end{equation}
where for the tilted ferromagnetic state the charge distribution is just a binomial of the form
\begin{equation}
    p_q = \binom{\ell_A}{q} s^{q}(1-s)^{\ell_A-q}, \hspace{0.5cm} s=\sin^2(\theta/2).
    \label{eq:binomial}
\end{equation}
Plugging the above into $H_{1/\alpha}(\{p_q\})$, we recover the expression already found in Ref. \cite{ares2023entanglement}, but now evaluated in $1/\alpha$
\begin{equation}
    H_{1/\alpha}(\{p_q\}) =\frac{\alpha}{\alpha-1}  \log \sum_q p_q^{1/\alpha} = \frac{\alpha}{\alpha-1}  \log  \sum_q\left[\binom{\ell_A}{q}\cos^{2(\ell_A-q)}(\theta/2)\sin^{2q}(\theta/2)\right]^{1/\alpha}.
\end{equation}
It is possible to evaluate the derivative with respect to the tilting angle in $[0,\pi/2]$, to check that indeed the correct ordering is satisfied. Some simple algebra yields
\begin{equation}
    \partial_\theta H_{1/\alpha}(\{p_q\})= \frac{2}{(\alpha-1) \sin \theta} \frac{\sum_q (q-\ell_A s) p_q^{1/\alpha}}{\sum_q p_q^{1/\alpha}}.
\end{equation}
In general, the above expression does not have a fixed sign and hence a given monotonicity of both $\Delta S^{(\alpha)}$ and $S^{(\alpha)}(\rho||\rho_Q)$. However, asymptotic monotonicity emerges for large $\ell_A$, where the binomial can be approximated by a Gaussian distribution with mean $\ell_A s$ and standard deviation $\sigma =\sqrt{\ell_A s(1-s)}$. Performing this approximation (which involves introducing the standardized variable $z = \frac{q- \ell_A s}{\sigma}$, and turning the sum over it into an integral) gives 
\begin{equation}
\label{eq:sum_into_integral}
     \frac{\sum_q (q-\ell_A s) p_q^{1/\alpha}}{\sum_q p_q^{1/\alpha}} \simeq \frac{1-2s}{2} (\alpha-1),
\end{equation}
finally leading to the result 
\begin{equation}
     \partial_\theta H_{1/\alpha}(\{p_q\}) \simeq \frac{1-2\sin^2(\theta/2)}{\sin \theta} = \cot\theta,
\end{equation}
which is  $\alpha$-independent, and positive in the region $[0,\pi/2]$. Note that this is exactly what is obtained by taking the $\theta$ derivative of \eqref{eq:asymptotic_divergence_smallt}, confirming the asymptotic equivalence of the behaviour of the standard tilted ferromagnetic state and its Gaussianised version at large system size.
This confirms that all R\'enyi divergences and asymmetries correctly capture the fact that larger tilting angles below $\pi/2$ corresponds to larger symmetry breaking. 
Note however that this approximation breaks down in two regimes: $\alpha \ll \ell_A^{-1}$ (this would correspond to $\Delta S^{(\alpha \to \infty)}$, which is known to be non-monotonic in $\theta$ from Ref. \cite{ares2023entanglement}) and for $\alpha \gg \ell_A$. In the latter regime the Gaussian approximation fails as the factor $1/\alpha$ makes the binomial into a uniform distribution. However, monotonicity in $\theta$ persists as the limit $\alpha \to \infty$ is approached: in Eq. \eqref{eq:sum_into_integral}, the average of the charge $q$ is approximated by the midpoint of the charge interval, $\ell_A/2$, giving
\begin{equation}
    \frac{\sum_q (q-\ell_A s) p_q^{1/\alpha}}{\sum_q p_q^{1/\alpha}} \simeq \frac{\ell_A}{2}(1-2s) = \frac{\ell_A}{2}\cos\theta,
\end{equation}
which, plugged in the derivative, yields an $\alpha$-dependent expression
\begin{equation}
    \partial_\theta H_{1/\alpha}(\{p_q\}) \simeq  \frac{\ell_A}{2} \frac{\cot \theta}{\alpha-1}.
\end{equation}
This function is still larger than zero in the relevant interval $[0,\pi/2]$, but with a damping factor of order $1/\alpha$ which makes the derivative vanish in the limit $\alpha \to \infty$. This reflects the fact that, in this limit, $H_0(\{p_q\})=\log (\ell_A+1)$, regardless of the tilting angle of the initial state. 
Hence we conclude that the tilted ferromagnetic state satisfies, schematically,
\begin{equation}
    S^{(\alpha)}(\theta_2) > S^{(\alpha)}(\theta_1) \iff \theta_2 > \theta_1
    \hspace{0.5cm} \forall\, \alpha \in [1/2,\infty),
\end{equation}
that is, at $t=0$ a larger tilt always means a larger asymmetry, whatever the value of $\alpha$.

\subsection{R\'enyi divergences for large times in free fermion models}
\label{subsec:latetimes}
On the other hand, here we show that the large-time regime can be studied for any $\alpha$ by adapting the approach of Ref. \cite{rylands2024microscopic}.
In section \ref{sec:replicatrick}, we have shown that a naive Taylor expansion of the charged moments leads to an asymptotic result at large times, Eq. \eqref{eq:divergencetlarge}, which cannot be extended to all values of $\alpha$. This is the case also for the situation studied in \cite{murciano2024entanglement}, where the asymptotic behaviour of $\Delta S^{(\alpha)}$ does not admit a continuation to $\alpha =1$. In this appendix, we adapt (with some modifications) the argument of Ref. \cite{rylands2024microscopic} to the evaluation of the R\'enyi divergences and use it to extract the large time behaviour in full generality. We consider equation \eqref{eq:Smn_freequench},
\begin{equation}
    S^{(m,n)}(\rho_A||\rho_{A,Q}) = n \log \int \prod_{k=1}^m \frac{d\lambda_k}{2\pi}  \exp\left\{\sum_{k=0}^{m}\int \frac{dk}{4\pi}\ell_k(t) f(\lambda_{k+1,k}) \right\},
\end{equation}
where we have neglected the initial state contribution which eventually disappears for $n(m+1)=1$.
We begin by observing that this can be expressed in terms of two functions, $g(k) = \max(\vartheta(k) ,1-\vartheta(k)) $ and $r(k) = \frac{\min(\vartheta(k) ,1-\vartheta(k))}{\max(\vartheta(k) ,1-\vartheta(k))}$  as
\begin{equation}
     S^{(m,n)}(\rho_A||\rho_{A,Q}) = n \log \int \prod_{k=1}^m \frac{d\lambda_k}{2\pi}  \exp\left\{\sum_{k=0}^{m}\int \frac{dk}{4\pi}\ell_k(t) \left(\log g(k) + \log(1+r(k)e^{-2i\sigma_k\lambda_{j+1,j}})\right) \right\}, 
\end{equation}
where $\sigma_k = \text{sign}(1-2\vartheta(k))$. This second form can then be simplified by introducing an expansion in terms of the modes
\begin{equation}
    J_q = \int_{-\pi}^\pi \frac{d\varphi}{2\pi} e^{i\varphi q} \exp{ \int \frac{dk}{4\pi} \ell_k(t)\left(\log g(k) + \log(1+r(k)e^{-2i\sigma_k\varphi})\right) },
    \label{eq:Jq}
\end{equation}
which satisfy $\sum_q J_q=1$. The coefficients $J_q$ have a direct physical meaning, as they encode the effective charge distribution associated with quasiparticle pairs whose two constituents remain inside $A$ at time $t$. Pairs sharing a partner with the complement are in a charge-diagonal state, and they contribute
classical charge noise to $Q_A$ but no asymmetry, so $J_q$ is the charge distribution
carried by the pure part of $\rho_A$\footnote{As discussed in Section \ref{sec:replicatrick}, this statement has to be interpreted at the hydrodynamic ballistic scale, where ballistic transport separates the RDM in two parts which are rigidly separated, one pure and one mixed, such that the pure contains the asymmetry. See for example the discussion of \cite{rottoli2024entanglementhamiltoniansquasiparticlepicture,travaglino2024}. Beyond the ballistic scale, the $J_q$ are not guaranteed to be positive.} and not the full counting statistics of $Q_A$. In terms of such objects,
 \begin{equation}
       S^{(m,n)}(\rho_A||\rho_{A,Q}) = n \log \sum_q J_q^{m+1},
\end{equation}
and therefore, in the physical limit \eqref{eq:physical_limit}:
\begin{equation}
\label{eq:renyi_divergence_moments_J}
S^{(\alpha)}(\rho_A||\rho_{A,Q}) = \frac{\alpha}{\alpha-1} \log \sum_q J_q^{1/\alpha}.
 \end{equation}
Crucially, this is the classical Rényi entropy of the distribution ${J_q}$ which indeed is positive in the hydrodynamics regime we are considering. The index
$\alpha$ then selects which fluctuations are probed, as discussed in appendix \ref{app:bulk_tail}: $\alpha<1$ favours bulk fluctuations, while increasing $\alpha>1$ weights more and more the rare charge sectors in the tails.
Interestingly, the expression in terms of $J_q$ allows one to take the physical limit immediately, without incurring in the problems which are present when sticking to the integral representation. Moreover, comparing Eq. \eqref{eq:renyi_divergence_moments_J} with the expression for $\Delta S^{(\alpha)}$ \cite{rylands2024microscopic}, it is clear that the mapping $S^{(\alpha)} \simeq \Delta S^{(1/\alpha)}$ holds. Since the mode decomposition is valid at any time, we conclude that for free fermionic systems in quenches from squeezed states the relation \eqref{eq:alphatooneoveralpha} holds at all times in the quasiparticle regime. 

At this point, the large-$t$ limit can be taken by expanding the $J_q$. As in the main text, this can be done by noting that $\ell_k(t)$ is a small parameter in the expansion of the exponential. Together with the observation that $r_k<1,$ which allows to expand in Taylor series the second logarithm, this allows to perform the simplifications
\begin{equation}
    \begin{split}
        J_q &\simeq \int_{-\pi}^\pi \frac{d\varphi}{2\pi} e^{i\varphi q} \left(1+ \int \frac{dk}{4\pi} \ell_k(t)\left(\log g(k) + \log(1+r(k)e^{-2i\sigma_k\varphi})\right) \right) \\&= \delta_{q,0} (1-G(t)) +   \int \frac{dk}{4\pi} \ell_k(t) \sum_{s=1}^\infty r(k)^s \frac{(-1)^{s+1}}{s}  \int_{-\pi}^\pi \frac{d\varphi}{2\pi} e^{i\varphi q} e^{-2is\sigma_k \varphi},
    \end{split}
\end{equation}
where we have introduced $G(t) = -\int\frac{dk}{4\pi} \ell_k(t) \log g(k)>0$. 
The presence of the factor $\sigma_k$ implies that for $\vartheta(k)<1/2$ the integral over $\varphi$ produces a delta function restricting to negative values of $s$, while for the opposite regime only positive values of $s$ appear. In both cases, it is only even values which enter the evaluation. We therefore introduce the auxiliary quantities
\begin{equation}
    \Lambda_q^{\pm} = \int_{\vartheta(k)\gtrless\frac{1}{2}} \frac{dk}{4\pi} \ell_k(t) r(k)^q ,\label{eq:Lambdapm}
\end{equation}
which finally allows to obtain the leading order expression for the modes,
\begin{equation}
    J_{2q} = \delta_{q,0} (1-G(t))  + \Theta(q) \frac{(-1)^{q+1}}{q}\Lambda_q^+  + \Theta(-q) \frac{(-1)^{q+1}}{q}\Lambda_q^-  .
\end{equation}
In fact, it is easy to show that the dominant terms are $G(t)$ and $\Lambda_1^{\pm}$. The former is known to scale as $t^{-3}$ for large times as it is the same quantity appearing for example in Ref. \cite{murciano2024entanglement}. Moreover, observing that $\log(1+r(k)) = -\log g(k)$, it is immediate to observe that $G(t) = \Lambda_1^+ + \Lambda_1^-  + \dots$ and therefore the two share this scaling\footnote{In principle, one can expand \eqref{eq:Lambdapm} around the slow modes which occur at momentum $k=0,\pi$ and obtain the scaling exactly. This is identical to calculations performed in detail in Ref. \cite{murciano2024entanglement}, and we therefore refer the reader to such work for more details. Performing this expansion of $\Lambda_q^{\pm}$ properly leads to the conclusion that $ \Lambda_q^{\pm} \sim t^{-1-2q}$ which confirms the $q=1 $ claim.}. Hence we conclude that the Rényi divergence admits the large-t approximation
\begin{equation}
\begin{split}
      S^{(\alpha)}(\rho_A||\rho_{A,Q}) &\simeq \frac{\alpha}{\alpha-1} \log \sum_q J_{2q}^{1/\alpha} \simeq \frac{\alpha}{\alpha-1} \log \left( (1-G(t))^{1/\alpha} + (\Lambda_1^+)^{1/\alpha} + (\Lambda_1^-)^{1/\alpha}\right) \\
    &\simeq  \frac{\alpha}{\alpha-1} \log \left( 1-\frac{1}{\alpha}G(t) + (\Lambda_1^+)^{1/\alpha} + (\Lambda_1^-)^{1/\alpha}\right),
\end{split}
\end{equation}
which is perfectly well-defined in all the range of values of $\alpha$. In particular, depending on the regime only one of the two terms will dominate. For $\alpha <1$, $G(t)\gg (\Lambda_1^{\pm})^{1/\alpha}$ and therefore the leading contribution is 
\begin{equation}
     S^{(\alpha<1)}(\rho_A||\rho_{A,Q}) = -\frac{1}{\alpha -1} G(t) =  \frac{1}{\alpha-1 } \int \frac{dk}{4\pi} \ell_k(t) \log(g(k)),
\end{equation}
which confirms the result of the main text \eqref{eq:divergencetlarge}.
For tilted ferromagnetic initial state, expanding the integral gives 
\begin{equation}
    S^{(\alpha<1)}(\rho_A||\rho_{A,Q}) = \frac{C_\alpha}{t^3} \left(1+ 8\frac{\cos^2\theta}{\sin^4\theta}\right),
\end{equation}
where $C_\alpha$ is an $\alpha$- and $\ell$-dependent constant. Importantly, the function of $\theta$ is strictly decreasing in $[0,\pi/2]$, independently on the value of $\alpha$.

On the other hand, for $\alpha >1$ the relation is inverted, $G(t)\ll (\Lambda_1^{\pm})^{1/\alpha}$ and it is therefore the second term which contributes the most to the logarithm. This gives the remarkable scaling
\begin{equation}
      S^{(\alpha>1)}(\rho_A||\rho_{A,Q}) = \frac{\alpha}{\alpha -1} \left((\Lambda_1^+)^{1/\alpha} + (\Lambda_1^-)^{1/\alpha}\right) \sim \frac{1}{t^{3/\alpha}}.
\end{equation}
Note that by \eqref{eq:alphatooneoveralpha} this describes the approach to infinity of both the Rényi divergence for $\alpha>1$ and the difference of Rényi entropies with $\alpha<1$, in which case the scaling would be $t^{-3\alpha}$. In the latter case, this dependence on $\alpha$ can be expected, as for $\alpha \to 0$ the Rényi entropies simply reproduce the support of the states, which is independent of the tilting. 
We can also assess the dependence of $S^{(\alpha)}(\rho_A \Vert \rho_{A,Q})$ on the initial tilting angle as $t\to \infty$. Indeed, $\Lambda_1^+$ is determined by an  expansion around the slow mode $k=\pi$. Since in this region we have:
\begin{equation}
    r(k\sim\pi) = \frac{1-\cos \Delta_k}{1+\cos \Delta_k} \simeq \frac{1}{4} \tan^4(\theta/2) (k-\pi)^2,
\end{equation}
the residual dependence on the tilting angle remains only in this term. On the other hand, $\Lambda_1^-$ involves $\frac{1+\cos \Delta_k}{1-\cos \Delta_k}$, and therefore the expansion produces a factor  $\frac{1}{4} \cot^4(\theta/2) k^2$ (note that the two terms $k^2$ and $(k-\pi)^2$ are precisely what give the scaling $t^{-3}$ once the residual integral in $k$ is performed in the small shell of size $\ell/2t$ around the slow modes). Hence the full angle dependence can be encoded as
\begin{equation}
    S^{(\alpha>1) } \simeq \frac{\tilde C_\alpha}{t^{3/\alpha}}\left(\tan(\theta/2)^{4/\alpha}+\cot(\theta/2) ^{4/\alpha}\right) =  \frac{2\tilde C_\alpha}{t^{3/\alpha}}\cosh\left(\frac{4}{\alpha} \log \cot(\theta/2)\right),\label{finallarget}
\end{equation}
where $\tilde C_\alpha$ is another $\alpha$- and $\ell$-dependent constant, different from the one above. The second function in \eqref{finallarget} is again monotonically decreasing for $\theta \in [0,\pi/2]$ for all values of $\alpha$. 

Finally, the relative entropy limit introduces a logarithmic correction to the scaling, of the form $t^{-3}\log t$, as discussed in \cite{rylands2024microscopic}: in our notation, the limit $\alpha \to 1$ gives from de l'Hopital rule
\begin{equation}
    S(\rho_A||\rho_{A,Q} )\simeq - \Lambda_1^+ \log \Lambda_1^+ - \Lambda_1^- \log \Lambda_1^- \sim \frac{2\tilde C_{1} \log t}{t^{3} }\cosh\left(4 \log \cot(\theta/2)\right),
\end{equation}
which confirms the scaling and also that the angle ordering is the same as the previous cases.

This confirms that in all regimes of $\alpha$, the ordering of $S^{(\alpha)}(\rho_A\Vert \rho_{A,Q})$ as a function of $\theta$ is reversed,
\begin{equation}
    S_{t\gg \ell}^{(\alpha)}(\theta_2) < S_{t\gg \ell}^{(\alpha)}(\theta_1) \iff \theta_2 > \theta_1
    \hspace{0.5cm} \forall\, \alpha \in [1/2,\infty), 
\end{equation}
which signals an $\alpha$-independent Mpemba effect.
As clear from \eqref{finallarget}, for $\alpha>1$ the late-time decay law still depends on $\alpha$, but the Rényi index plays the role of a damping factor without changing the ordering of the curves.
This is a remarkable fact: the existence of the crossing does not depend on which of the sandwiched R\'enyi divergences is used to quantify the asymmetry (the crossing time, on the other hand, may depend on $\alpha$, as the power law determining the large-time behaviour).

To understand the meaning of this result, we can consider the physical meaning of $J_q$.
In the quasiparticle picture, it is the probability distribution of the charge carried by the pairs that are fully contained in $A$ at time $t$, measured with respect to its average value (which is constant and equal to the average charge of the initial state).
These pairs form the coherent, pure component of $\rho_A$: each of them is in a superposition of two charge states, and it is this superposition that breaks the symmetry.
The pairs sharing a partner with the complement are instead in a charge-diagonal mixed state, and do not contribute to the asymmetry. One can interpret $J_q$ as the distribution of the charge fluctuations responsible for the symmetry breaking: it depends on time through $\ell_k(t)$, and its width is a measure of how delocalised the coherent part of the state is over different charge sectors.

This identification makes the role of $\alpha$ transparent, since the R\'enyi index $1/\alpha$ selects which charge fluctuations are probed.
For $\alpha=\tfrac12$ the entropy is the collision entropy, $-\log\sum_q J_q^2$, which is controlled by the bulk of the distribution and is directly related to the fidelity between $\rho_A$ and its symmetrised version.
For $\alpha=1$ one obtains the Shannon entropy of $J_q$, i.e.\ the relative entropy, which governs the typical charge fluctuations and sets the error exponent for distinguishing $\rho_A$ from $\rho_{A,Q}$ on many copies \cite{cover2006elements}.
For $\alpha>1$ the index $1/\alpha<1$ gives increasing weight to the rare charge sectors in the tails of $J_q$, which are those that control the strong-converse exponent of the same discrimination task.
The three regimes therefore probe the bulk, the typical part and the tails of the distribution of the coherent charge, which are different physical aspects of the same state.

The limit $\alpha\to\infty$ is excluded from the statement above because it is trivial.
For $\alpha\to\infty$ the divergence becomes the max-relative entropy, $\log\min\{\lambda:\rho_A\le\lambda\,\rho_{A,Q}\}$, which in terms of $J_q$ is the Hartley entropy
\begin{equation}
    S^{(\infty)}(\rho_A\|\rho_{A,Q}) = H_0\big(\{J_q\}\big)=\log\,\#\{q:\,J_q>0\}.
\end{equation}
This quantity only counts how many charge sectors are populated, and it is blind to their weights.
In the tilted ferromagnet every pair with momentum $k\neq0,\pi$ has a non-trivial occupation, $0<n(k)<1$, for any $0<\theta<\pi$.
The number of populated sectors is then fixed by the number of pairs fully contained in $A$, and it coincides for all tilting angles. 
There is no swap because there is nothing to swap: the curves for different tilt angles coincide at all times, and the scaling $t^{-3/\alpha}$ implies that the approach to zero has a different law for $\alpha \to \infty$.
Only the symmetric limit $\theta=0,\pi$, in which all occupations are $0$ or $1$ and the state has no asymmetry, is different. 

In this language, the Mpemba effect is a statement about the time evolution of the distribution $J_q$: the more tilted state starts with a broader distribution of the coherent charge and, at late times, ends up with a narrower one.
Since this holds for every $\alpha\in[1/2,\infty)$, the inversion is independent of how the charge fluctuations of the coherent part are weighted, whether by their typical values, their mean, or their extremes. 
This is a much stronger statement than the crossing of the relative entropy alone, which could be produced by a mere exchange of weight between the bulk and the tails of $J_q$.
The inversion is therefore more stringently related to the true asymmetry carried by the subsystem, and not to the measure used to quantify it.

\section{An ordered SWAP model}
\label{sec:swap_model}

In this section we introduce a simple toy model that captures the minimal ingredients of symmetry restoration, inspired by constructions analogous to those of the Page model for black hole evaporation in  Refs.~\cite{page1993average,page1993information,Ares2024BlackHole}. 
Remarkably, we find that the Mpemba effect is robust even in this model.
The only exceptions occur for some initial states for $\alpha<1/2$, where the R\'enyi divergence is no longer guaranteed to be an asymmetry monotone.
We emphasize, however, that we do not expect such robustness to be a universal feature of the Mpemba effect. On the contrary, we find it particularly striking that the effect persists across the entire family of asymmetry monotones in all the physically relevant cases considered here, despite the different properties of the measures involved.

The toy model we introduce is constructed so that $\rho_A(t)$ remains pure throughout the time evolution, thereby circumventing the difficulties associated with evaluating $S^{(\alpha)}(\rho_A||\rho_{A,Q})$ for mixed states.
Within this framework, we investigate three classes of initial states.
For the first class, we find that the Mpemba effect is absent for all values of $\alpha$, including $\alpha=1$, which corresponds to the standard entanglement asymmetry.
For the second class, instead, the Mpemba effect occurs for the entire family of asymmetry measures, providing an example of fully robust behaviour.
Finally, we identify a third class of initial states for which the occurrence of the Mpemba effect depends on the chosen measure: it is observed for values of $\alpha>1/2$, but disappears for $\alpha<1/2$, when the R\'enyi divergence ceases however to be an asymmetry monotone.

The toy model is defined as follows. We consider a state $\ket{\psi_0}=\ket{\psi_0}_A \otimes \ket{\psi_0}_B$ and we are interested in the evolution of the $\alpha$-Rényi divergence of $\rho_A(t)$. Let
\begin{equation}
\ket{\psi_0}_A = \otimes_{j=1}^N(\sqrt{1-p_j}\ket{0}_j + \sqrt{p_j}\ket{1}_j)\,,
\end{equation}
where $\ket{0}$ has charge 0 and $\ket{1}$ has charge +1. Let also $\ket{\psi_0}_B=\ket{0}^{\otimes |B|}$, where the size of subsystem $B$ is $|B|\ge |A|\equiv N$. At any discrete time $t\ge 0$, our protocol consists in orderly swapping the $t$-th qubit in $A$ with the corresponding qubit in $B$, so $\rho_A(t)$ is progressively symmetrized while remaining pure:
\begin{equation}
\label{eq:SWAP_hom_time_evolved_state}
\ket{\psi(t)}_A = \otimes_{j=1}^t \ket{0}_j\otimes_{j=t+1}^N(\sqrt{1-p_j}\ket{0}_j + \sqrt{p_j}\ket{1}_j)\,.
\end{equation}
The process stops at time $t=N$, and the final state $\ket{\psi(t=N)}_A=\ket{0}^{\otimes N}$ has zero asymmetry. This time-evolution protocol is similar in spirit to the one devised by Page in order to solve the black-hole information paradox \cite{page1993information}: in Page's model, qubits are progressively emitted (at a linear rate) from one subsystem of a \textit{random} pure state, becoming part of the complementary subsystem, whereas in our case the initial state $\ket{\psi_0}$ is bipartite and not random, with asymmetry specified by its total charge distribution. To make a connection with the quasiparticle picture of fermionic quenches, we observe that orderly swapping qubits between region $A$ and its complement is an extremely simplified model of ballistic propagation, in which the qubits further from the boundary are the last to \lq\lq escape \rq\rq the system, that is, they are tantamount to the slow modes. As we show below, this model captures their crucial role in determining the presence of Mpemba effect.

Since $\rho_A(t)$ is pure, at any time $t\in \{0,\dots,N\}$, $S^{(\alpha)}_A(t)= H_{1/\alpha}(\{p_{q}(t)\})$, where $p_q(t)$ are the evolved charge probabilities, with $q$ eigenvalue of $Q_A$. The qubits carrying charge asymmetry at $t$ are those at sites $t < j \le N$, and at time $t$ the total charge $Q_A$ can take values $q=0,\dots, N-t$. The charge probabilities are given by
\begin{equation}
p_q(t) = \sum_{\substack{S \subseteq \{t+1,\dots,N\} \\  |S|=q}} \,\prod_{j \in S} p_j \prod_{j \notin S} (1-p_j)\,.
\end{equation}

\begin{figure}[t]
    \centering
    \includegraphics[width=0.49\linewidth]{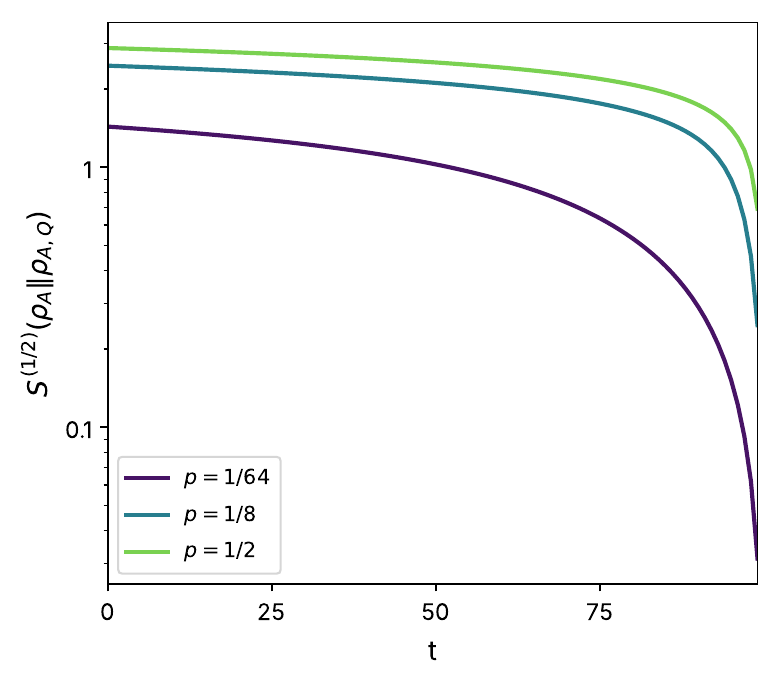} \includegraphics[width=0.49\linewidth]{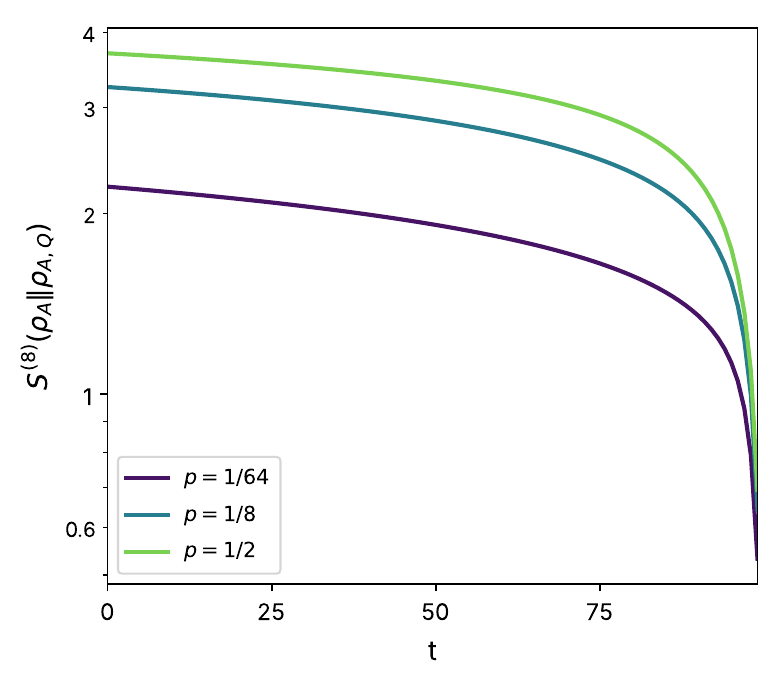}
    \caption{Rényi divergences $S_A^{(\alpha)}(\rho_A||\rho_{A,Q})$ for the ordered SWAP model, starting from a homogeneous state. At time $t$, the state in subsystem $A$ is described by Eq. \eqref{eq:SWAP_hom_time_evolved_state}, with $p_j=p$ for all $j$. The subsystem size is $|A|=N=100$. \textit{Left}: Rényi index $\alpha=1/2$. \textit{Right}: Rényi index $\alpha=8$. There is no Mpemba effect at any value of $\alpha\ge 1/2$, the range in which $S_A^{(\alpha)}$ is a resource-theoretic monotone.}
    \label{fig:swap_hom}
\end{figure}

Let us first consider the simple case of a homogeneous initial state with all $p_j=p$. In this case, the probability distribution is binomial
\begin{equation}
p_q(t) = \binom{N-t}{q}p^q(1-p)^{N-t-q}\,.
\end{equation}
For this initial state, it is easy to show analytically that there is no Mpemba effect in $S^{(1/2)}_A(t)= H_2(\{p_q(t)\})= \Delta S_A^{(2)}(t)$, which can be written as
\begin{align}
\label{eq:swap_H_2}
H_2(\{p_q(t)\})&= -\log \sum_{q=0}^{N-t} p^2_q(t)\nonumber \\ &=-\log \int_0^{2\pi}\frac{d\beta}{2\pi} [1-4p(1-p)\sin^2(\beta/2)]^{N-t}\,.
\end{align}
The integral is monotonically decreasing in $p \in [0,1/2]$ and so the asymmetry is an increasing function of $p$ at all times, i.e. the ordering never changes as time passes. The absence of Mpemba effect can be proved analytically also in the cases $\alpha=1$, where it follows from the Shepp-Olkin theorem on the concavity of the Shannon entropy (see \cite{shepp1981entropy,hillion2017proof} and the supplemental material of \cite{Mazzoni2025asymmetry}), and trivally for $\alpha \to \infty$, where the Rényi divergences of all probability distributions collapse to the same value. For generic values of $\alpha\ge 1/2$, we numerically certify that there are no crossings in $S^{(\alpha)}_A(t)$, see Fig. \ref{fig:swap_hom}. Thus, in this protocol, there cannot be Mpemba effect when the initial state is homogeneous. 

\begin{figure}[t]
    \centering
    \includegraphics[width=0.49\linewidth]{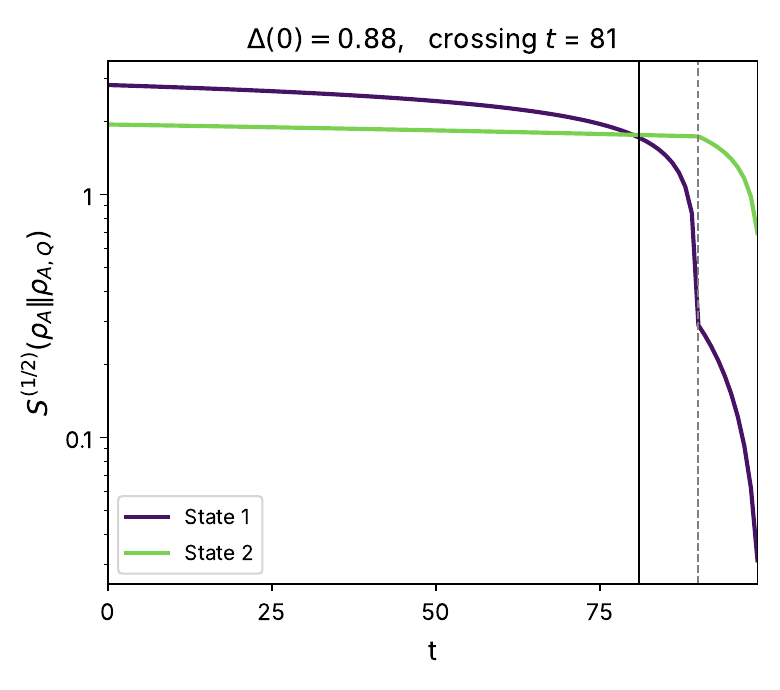} \includegraphics[width=0.49\linewidth]{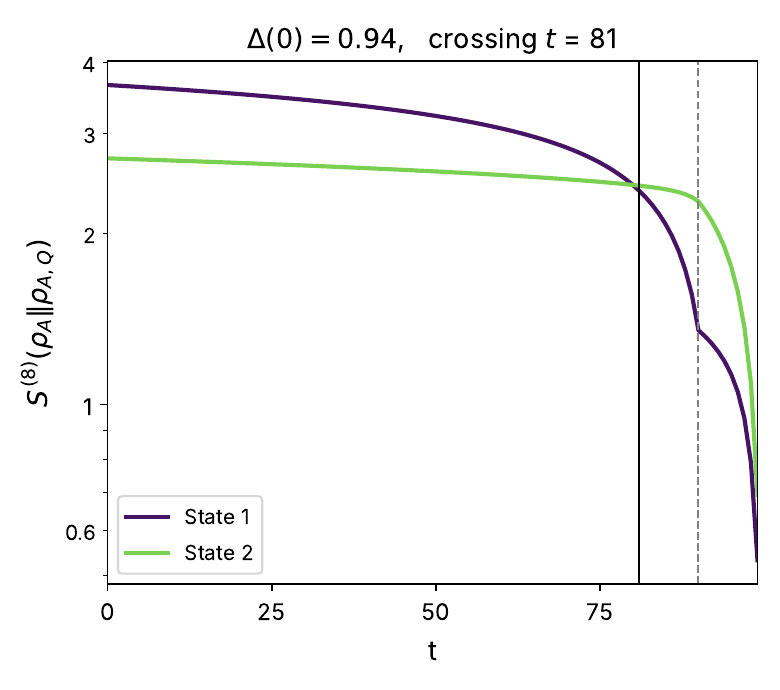}
    \caption{Rényi divergences $S_A^{(\alpha)}(\rho_A||\rho_{A,Q})$ for the ordered SWAP model, starting from the inhomogeneous state \eqref{eq:swap_inhom_state_initial}. In both figures, the total subsystem size is $|A|=N=100$, and we take $N_F=90$, $N_S=10$. \textit{Left}: Rényi index $\alpha=1/2$. \textit{Right}: Rényi index $\alpha=8$. For both values of $\alpha$, we consider two initial state: state 1 has $(p_{(F)},p_{(S)})=(1/2,1/64)$ and state 2 has $(p_{(F)},p_{(S)})=(1/64,1/2)$. The quantity $\Delta(0)$ is the initial difference in the $\alpha$-Rényi divergence of the two states. Because in state 1 most of the asymmetry is stored in the fast qubits, which are much more than the slow qubits, its initial asymmetry is larger than in state 2, as is the rate at which it decreases before $t=N_F$, while the situation is reversed for $t> N_F$. This explains the presence of Mpemba effect, which we observed in $S_A^{(\alpha)}$ for all other values of $\alpha \ge 1/2$.}
    \label{fig:swap_inhom}
\end{figure}

The second simple class of initial states consists of states with inhomogeneous tilting, such that the tilting is homogeneous in the first $N_F$ qubits, the \textit{fast} qubits, and in the remaining $N_S$ qubits, the \textit{slow} qubits, with $N_F + N_S=N$. In subsystem $A$, the initial state is therefore
\begin{equation}
\label{eq:swap_inhom_state_initial}
\ket{\psi_0}_A = \left(\sqrt{1-p_{(F)}}\ket{0} + \sqrt{p_{(F)}}\ket{1}\right)^{\otimes N_F} \otimes \left(\sqrt{1-p_{(S)}}\ket{0} + \sqrt{p_{(S)}}\ket{1}\right)^{\otimes N_S}\,,
\end{equation}
where $p_{(F)}$ and $p_{(S)}$ parametrise the two tilting angles. The first tensor product is  over qubits $1,\dots,N_F$, and the second over qubits $N_F+1,\dots,N$. The dynamical protocol is the same as before, with qubits orderly swapped between subsystems $A$ and $B$. At time $t=1,\dots,N_F$, the $t$-th fast qubit is swapped with $\ket{0}$, so there are $N_F-t$ fast qubits left and $N_S$ slow qubits left. At $t=N_F+1,\dots,N$, the $t$-th qubit, which is now in the slow group, is swapped, leaving only $N-t$ slow qubits. At late times, because there are only slow qubits left, the charge probability distribution in $A$ is
\begin{equation}
p_q(t) = \binom{N-t}{q}p_{(S)}^q(1-p_{(S)})^{N-t-q}\,, \quad t\ge N_F\,.
\end{equation}
At early times, however, the probability distribution $p_q(t)$ is given by the convolution of the fast and slow qubit distributions. Indeed, the total charge in $A$ can be regarded as the sum of the two discrete random variables $Q_F$ and $Q_S$, i.e. the total charges of the fast and slow qubits in $A$, respectively. Therefore
\begin{align}
p_q(t) &\equiv {\rm P}(Q_A=q)=\sum_{j=0}^q{\rm P}(Q_F=j){\rm P}(Q_S=q-j)\nonumber \\
&=\sum_{j=0}^q\binom{N_F-t}{j}p_{(F)}^j(1-p_{(F)})^{N_F-t-j}\binom{N_S}{q-j}p_{(S)}^{q-j}(1-p_{(S)})^{N_S-q+j}\,, \quad t \le N_F\,.
\end{align}
As we show in Fig. \ref{fig:swap_inhom}, introducing a simple inhomogeneity in the initial state is enough to yield Mpemba effect, the onset of which depends on $N_F$, $N_S$, $q_{(F)}$ and $q_{(S)}$.

For the third class of initial states, we partition $A$ into a set of $N_F$ fast qubits and $N_S$ slow qubits, as before. To investigate the presence of Mpemba effect, we choose state 1 as in \eqref{eq:swap_inhom_state_initial}, with $p_{(F)}=1/2$ and $p_{(S)}\equiv p$
\begin{equation}
\label{eq:swap_third_example_1}
\ket{\psi_0^{(1)}}_A = \left(\frac{\ket{0} + \ket{1}}{\sqrt{2}}\right)^{\otimes N_F} \otimes \left(\sqrt{1-p}\ket{0} + \sqrt{p}\ket{1}\right)^{\otimes N_S}\,,
\end{equation}
whereas in state 2 we let $p_{(F)}=p<1/2$ and, in the slow qubits, we stagger the tilting (we take $N_S$ even)
\begin{equation}
\label{eq:swap_third_example_2}
\ket{\psi_0^{(2)}}_A = \left(\sqrt{1-p}\ket{0} + \sqrt{p}\ket{1}\right)^{\otimes N_F} \otimes \left[(\sqrt{1-p}\ket{0} + \sqrt{p}\ket{1})(\sqrt{p}\ket{0} + \sqrt{1-p}\ket{1})\right]^{\otimes N_S/2}\,.
\end{equation}
As is evident from Eq. \eqref{eq:swap_H_2}, the (classical) Rényi entropy $H_2$ of a probability distribution $\{p_q\}$, for a homogeneous product state, depends only on the variance of the single qubit $\sqrt{1-p}\ket{0}+\sqrt{p}\ket{1}$, given by $p(1-p)$ and invariant when $p$ and $1-p$ are swapped. Thus, for the two states above, $S^{(1/2)}_A(t)$ is larger in state 1 for $t\le N_F$, and it coincides exactly for the two states when $t > N_F$, yielding a weak version of the Mpemba effect. On the other hand, for other values of $\alpha$, $S^{(\alpha)}_A(t)$ depends in general on the entire probability distribution. Thus, as shown in Fig. \ref{fig:swap_staggered} below, $S^{(\alpha)}_A(t)$ is in general smaller in state 1 at late times, meaning that there is Mpemba effect.
\begin{figure}[!t]
    \centering
    \includegraphics[width=0.49\linewidth]{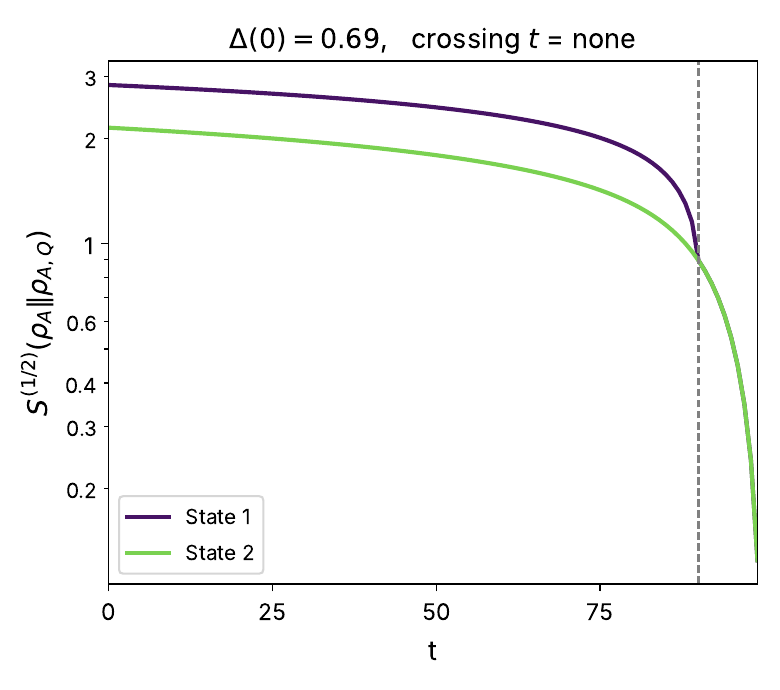} \includegraphics[width=0.49\linewidth]{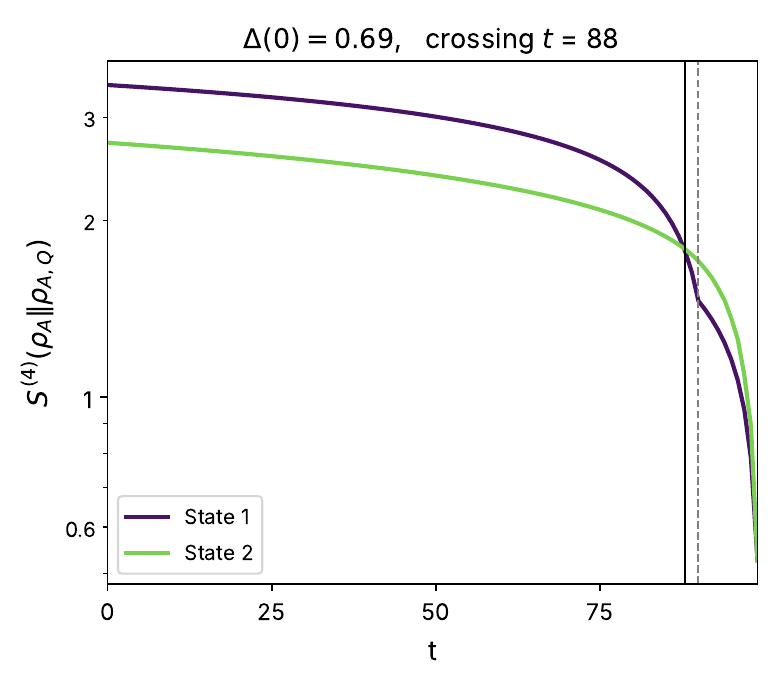}
    \includegraphics[width=0.49\linewidth]{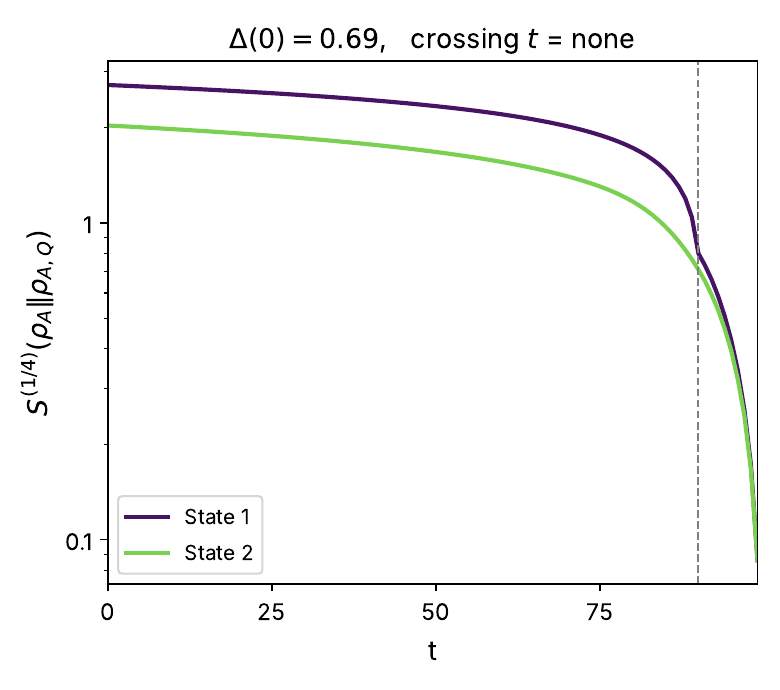} \includegraphics[width=0.49\linewidth]{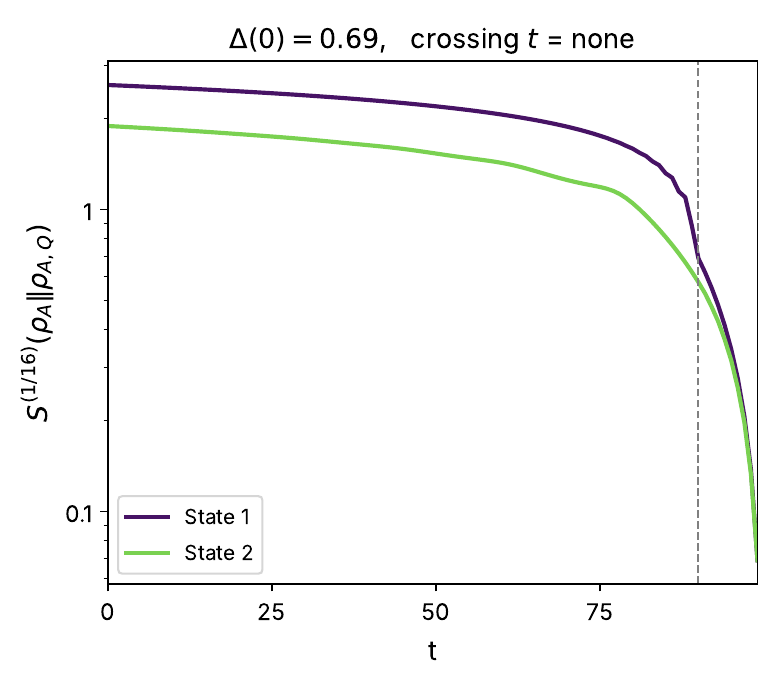}
    \caption{Rényi divergences $S_A^{(\alpha)}(\rho_A||\rho_{A,Q})$ for the ordered SWAP model, starting from the states \eqref{eq:swap_third_example_1} and \eqref{eq:swap_third_example_2}. In all panels, the total subsystem size is $|A|=N=100$, and $N_F=90$, $N_S=10$. 
    We fix $p=1/16$ for both states 1 and 2.
    $\Delta(0)$ is the initial difference in the $\alpha$-Rényi divergences.
    \textit{Top Left}: Rényi index $\alpha=1/2$. \textit{Top Right}:  $\alpha=4$. 
    \textit{Bottom Left}:  $\alpha=1/4$. \textit{Bottom Right}: $\alpha=1/16$.
    We numerically observe the  Mpemba effect for all  $\alpha >1/2$. For $\alpha=1/2$, the two curves coincide after a given time and for $\alpha<1/2$ there is no Mpemba effect.}
    \label{fig:swap_staggered}
\end{figure}

For $\alpha <1/2$, the $\alpha$-Rényi divergence is not a monotone in the resource-theoretic sense, however, for this particular dynamical protocol and with this choice of the initial states, $S_A^{(\alpha)}(t)$ decreases monotonically in time. In this case we do not observe Mpemba effect, signalling that here the value $\alpha=1/2$, which can be regarded as a \lq\lq degenerate case\rq\rq, separates probes which detect Mpemba effect and probes which do not. The states \eqref{eq:swap_third_example_1} and \eqref{eq:swap_third_example_2}, which are highly inhomogeneous, fall far from the cases discussed in the previous sections (both the quenches and the random unitary evolution), in which the initial state is homogeneous. The present example signals that it is possible to construct \textit{ad hoc} initial states and dynamical protocols in which, although all the asymmetry proxies correctly decrease throughout the symmetrization process, they do not display the same crossing phenomenology.

\section{When monotonicity matters}
\label{sec:matters}
In the previous sections, we have shown that the Rényi divergences and the difference of Rényi entropies display the same qualitative behaviour in several situations of interest, including pure states, homogeneous quenches, and Haar-random states. As discussed in Sec.~\ref{sec:replicatrick}, in homogeneous quench settings this agreement can be traced back to the structure of the degrees of freedom that contribute to the asymmetry. Up to corrections that are subleading on hydrodynamic scales, this contribution is effectively described by a pure state, for which the relation in Eq.~\eqref{eq:alphatooneoveralpha} holds exactly. This picture is further supported by the calculation presented in Sec.~\ref{sec:universality}, where both asymmetry quantifiers are expressed in terms of the charge fluctuations of the effective pure component of $A$, encoded in the distribution $\{J_q\}$. We have argued that this correspondence should hold quite generally at large scales, provided that the asymmetry is carried by the bulk of the subsystem and that tracing out its complement removes the relevant asymmetric contribution. Under these conditions, the hydrodynamic description establishes a direct relation between the two families of asymmetry quantifiers.

The correspondence, however, is not universal. In particular, the final example of the previous section demonstrates that introducing inhomogeneities in the initial state can give rise to a richer phenomenology (as also shown in \cite{oac-26}). Motivated by this observation, we now explore settings in which the assumptions underlying the hydrodynamic argument are deliberately relaxed. We present explicit examples where the Rényi entropy difference and the Rényi divergences exhibit qualitatively distinct behaviour. These examples reveal situations in which the monotonicity of the divergences captures information that is not directly encoded in the entropy-difference construction.

\begin{figure*}[t!] \centering \includegraphics[width=\linewidth]{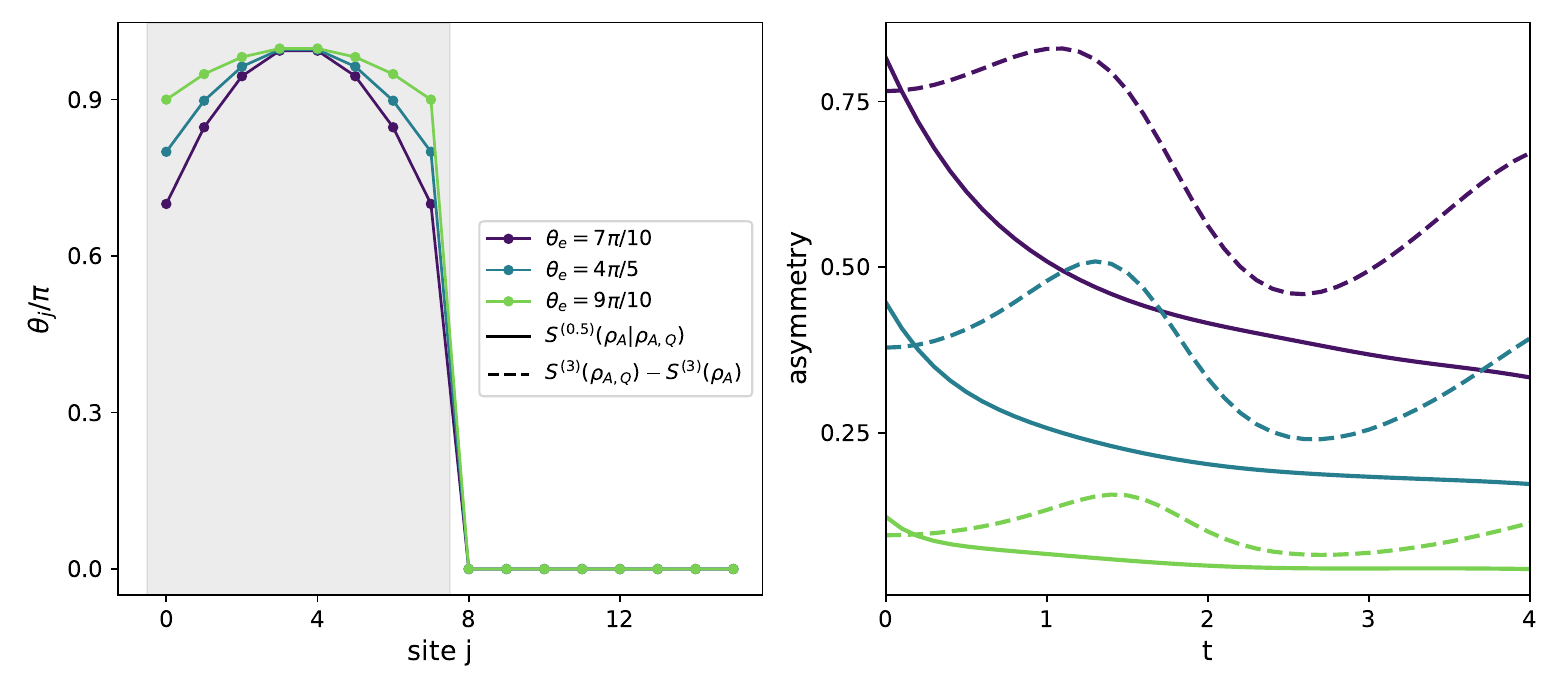} \caption{Dynamics of Rényi divergences and Rényi asymmetries following a quench from an inhomogeneous tilted-ferromagnetic state \eqref{eq:inhomogeneoustiltedferro} evolving under the hopping Hamiltonian Eq. \eqref{eq:hopping_hamiltonian}. Left panel: spatial profile of the tilting angle in the initial state; the complement $\overline{A}$ is initialized in the symmetric state with $\theta=0$, so the reduced dynamics of $A$ is symmetrised in the sense discussed in the text. Within the subsystem, the tilting is varied continuously with a sharp jump at the edges, where it takes the value $\vartheta_e$. Right panel: asymmetry dynamics. The Rényi divergences (thick lines) decrease from their initial values, consistently with Eq.~\eqref{eq:monotonicity-reduced-dynamics}. By contrast, the Rényi entropy differences (dashed lines) initially increase, and exhibit a nontrivial oscillation pattern.
The three pairs of curves correspond to the initial profiles on the left (same color scheme).} \label{fig:goodvsbadmonotones} \end{figure*}

\begin{figure}[t!] \centering \includegraphics[width=\linewidth]{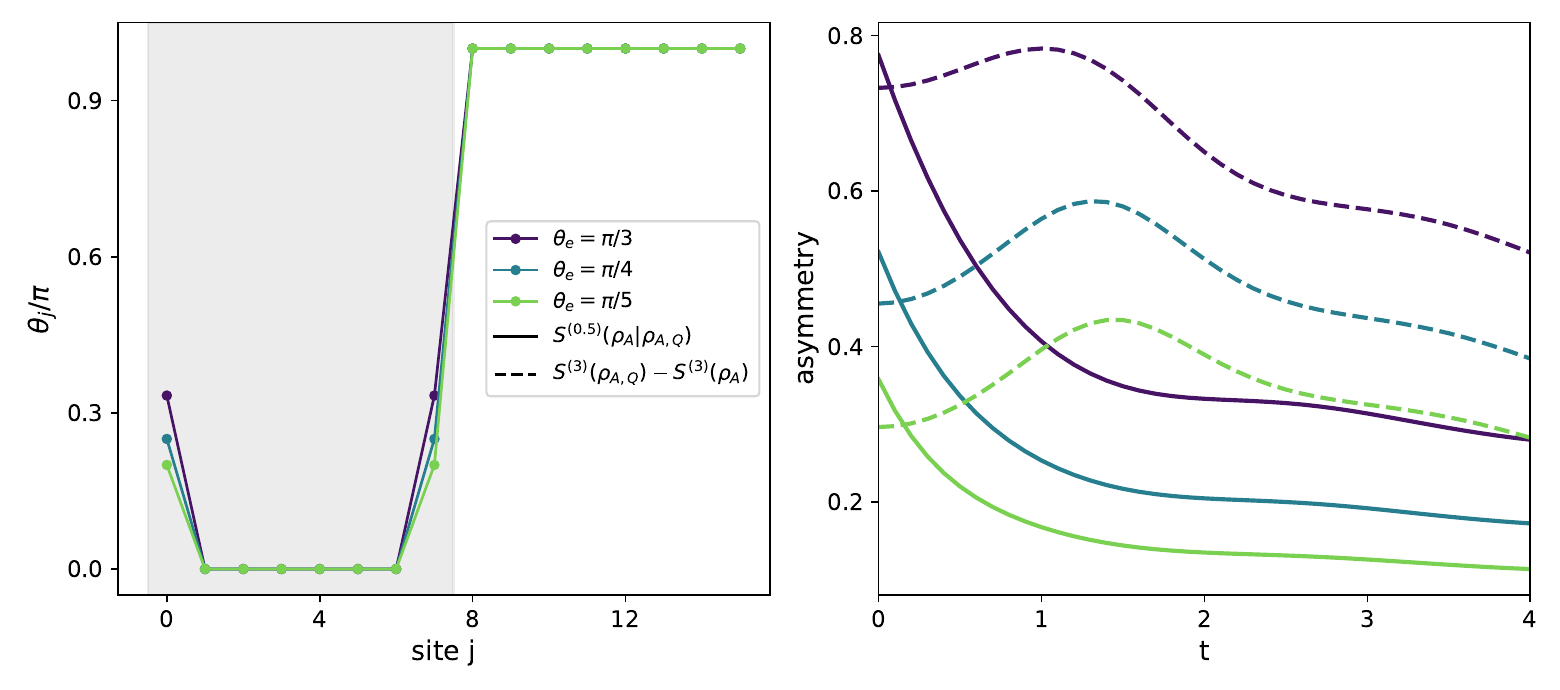} \caption{ Quench from an inhomogeneous tilted-ferromagnetic state evolving under the Heisenberg Hamiltonian, Eq. \eqref{eq:heisenberg-hamiltonian}. The spatial profile of the tilting angle is shown in the left panel. The complement $\overline{A}$ is initialized in the symmetric state with $\theta=\pi$. The bulk of the subsystem is symmetric and has the opposite ferromagnetic polarization, while symmetry breaking is localized at its edges. The Rényi divergences remain bounded by the initial value, as required for the resulting reduced dynamics, whereas the Rényi entropy difference displays a transient increase. This interacting example confirms that the distinction is not specific to free dynamics. The three pairs of curves corresponds to the initial profiles on the left (same color scheme).} \label{fig:heisenberg} \end{figure}

To this end, we study quenches from inhomogeneous tilted-ferromagnetic states, in which the tilting angle varies spatially along the chain,
\begin{equation}
    \ket{\psi} = \exp\left\{\frac{i}{2} \sum_j \theta_j \sigma_j^x\right\} \ket{\uparrow \uparrow \dots \uparrow}. \label{eq:inhomogeneoustiltedferro}
\end{equation}
Representative profiles (periodic boundary conditions are assumed) are shown in Figs.~\ref{fig:goodvsbadmonotones} and \ref{fig:heisenberg} for a free and an interacting model respectively. We choose the initial state of the complement $\overline{A}$ to be fully symmetric. Since the global Hamiltonian preserves the symmetry, and the symmetric complement cannot supply asymmetry to $A$ more than what is already within the subsystem of interest at the initial time,  any asymmetry monotone $\mathcal{A}$ should satisfy\footnote{For the R\'enyi divergence this is easy to prove. Let \(\mathcal G\) denote the symmetry twirling operation and write \(\rho_{AB}(0)=\rho_A(0)\otimes\rho_B(0)\), with \(\rho_B(0)\) symmetric.
For every \(\alpha\geq1/2\), data processing and additivity give
\[
\begin{aligned}
S^{(\alpha)}(\rho_A(t)\Vert\mathcal G_A(\rho_A(t)))
&\leq
S^{(\alpha)}(\rho_{AB}(t)\Vert\mathcal G_{AB}(\rho_{AB}(t)))
=
S^{(\alpha)}(\rho_A(0)\Vert\mathcal G_A(\rho_A(0))).
\end{aligned}
\].}
\begin{equation} \mathcal{A}\bigl(\rho_A(t)\bigr) \leq \mathcal{A}\bigl(\rho_A(0)\bigr) \qquad \text{for all }t. \label{eq:monotonicity-reduced-dynamics} \end{equation} 
More generally, from a physical perspective, one might expect a meaningful measure of symmetry breaking to decrease monotonically during symmetry restoration. However, such monotonicity cannot be imposed as a mathematical requirement, since unitary time evolution is not guaranteed to act as a symmetrizing operation at the subsystem level.
In particular, in systems with ballistic transport, it is always expected that the dynamics of $A$ should also be symmetrising up to finite size effects (of the full system, not of $A$), meaning that an asymmetry quantifier should be monotonically decreasing. This expectation is justified as in the ballistic scenario the asymmetry within $A$ is only expected to exit the subsystem through quasiparticle transport, and cannot return inside unless finite-size effects take place (a counterexample to this feature can arise due to curved trajectories \cite{alba2019entanglement,bertini2024dynamics,travaglino2026space}, which are however absent in free systems).
This provides a direct setting in which the implications of monotonicity can be tested.

Figure~\ref{fig:goodvsbadmonotones} shows the resulting dynamics. As required by Eq.~\eqref{eq:monotonicity-reduced-dynamics}, during the evolution the Rényi divergences never exceed their initial values. The Rényi entropy difference, instead, displays an initial increase, \begin{equation} \Delta S_A^{(\alpha)}(t) > \Delta S_A^{(\alpha)}(0) \label{eq:renyi-increase} \end{equation} for a finite interval of time. In the present setup, Eq.~\eqref{eq:renyi-increase} cannot be interpreted as an increase in asymmetry in the resource-theoretic sense because of the arguments just presented. Rather, it shows that $\Delta S_A^{(\alpha)}$ remains sensitive to changes in the spectrum and charge-sector structure of the reduced state that need not follow the ordering imposed by symmetric operations. The origin of the difference can be understood from the structure of the initial state. The tilting profile varies abruptly near the boundaries of $A$, while it varies smoothly within the subsystem, eventually reaching the tilting angle $\theta=\pi$ in its central region. The symmetry-breaking contribution is therefore localized near the subsystem boundaries, precisely where the assumptions underlying the bulk hydrodynamic description are least effective. In particular, the asymmetry cannot be associated with a smoothly evolving, effectively pure contribution in the bulk of $A$. The mechanism responsible for the agreement found in homogeneous quenches is thus absent, and the distinct properties of the two quantifiers become visible.

The interacting example in Fig.~\ref{fig:heisenberg} exhibits the same qualitative phenomenon. Here, we consider time evolution under the spin-$1/2$ Heisenberg Hamiltonian: 
\begin{equation}
H_{\mathrm{XXX}} = J\sum_j \left( S_j^x S_{j+1}^x + S_j^y S_{j+1}^y + S_j^z S_{j+1}^z \right), \label{eq:heisenberg-hamiltonian} 
\end{equation} 
which preserves the total magnetization along the $z$ direction. Once again, the initial state is chosen so that the complement $\overline{A}$ is symmetric, this time with opposite tilting $\theta=\pi$, while the subsystem $A$ is in the $\theta=0$ state except for the boundary points of the interval. As a consequence, no proper asymmetry monotone in subsystem $A$ can increase compared to its value at $t=0$. Nevertheless, the Rényi entropy difference develops a value larger than its initial one, while the divergence-based quantity respects the monotonicity bound. The effect is therefore not tied to the free-fermion structure of the hopping model, but persists in the interacting dynamics of the Heisenberg chain.

These examples clarify the complementary roles played by the two families of asymmetry proxies. For homogeneous initial states, and in the ballistic hydrodynamic regime, they are closely related, allowing the Rényi entropy difference to provide efficient access to phenomenology that is also captured by asymmetry monotones. In strongly inhomogeneous configurations, particularly when symmetry breaking is localized near subsystem boundaries, this correspondence can fail. In such cases, monotonicity provides an additional criterion for distinguishing a genuine increase of asymmetry from a more general rearrangement of the spectral and charge-sector properties of the reduced state.

\section{Conclusions}
\label{sec:conclusions}
In this work, we have investigated entanglement asymmetry from a resource-theoretic perspective, focusing on Rényi generalizations that retain the monotonicity properties of the relative entropy of asymmetry. Our main result is the identification of a remarkable connection between these quantities, defined through sandwiched Rényi divergences, and the conventional Rényi entanglement asymmetries, constructed as differences of Rényi entropies. Although the latter are not generally monotones, we have shown that the two families exhibit the same leading scaling behaviour in several physically relevant settings, being related by the simple inversion of the Rényi index, \(\alpha\leftrightarrow 1/\alpha\). This correspondence, exact for pure states, extends to homogeneous quantum quenches, including interacting integrable systems, and to Haar-random states in the appropriate asymptotic regimes. We have provided a physical interpretation of this result in terms of the structure of the reduced density matrix: in the hydrodynamic limit, the degrees of freedom responsible for symmetry breaking are effectively described by a pure component, while the remaining mixed degrees of freedom are already symmetric. Consequently, the asymmetry quantifiers are governed by the charge fluctuations of this effective pure component, explaining the persistence of the pure-state relation even for mixed reduced density matrices.

A particularly striking consequence emerges in the dynamics of the quantum Mpemba effect. For quenches from tilted ferromagnetic states, we have shown that the reversal in the ordering of the asymmetry curves persists throughout the entire family of sandwiched Rényi asymmetry monotones, for all finite Rényi indices in their monotonicity range. This establishes a stronger form of robustness than the observation of the effect through a single asymmetry measure. Indeed, different Rényi indices probe different regions of the charge distribution, ranging from its typical fluctuations to increasingly rare events. The persistence of the Mpemba effect across these different probes indicates that the reversal is not an artifact of a particular quantifier, but reflects a robust feature of the underlying symmetry-restoration dynamics. At the same time, we have uncovered an index-dependent algebraic relaxation at long times, showing that different monotones can approach the symmetric state with different asymptotic exponents, even though they agree on the occurrence of the Mpemba effect.

The correspondence between conventional Rényi asymmetries and genuine monotones is, however, not universal. We have illustrated the possible dependence of Mpemba crossings on the Rényi index through a simple solvable model, and identified more pronounced differences in quenches from spatially inhomogeneous initial states. In these situations, the conventional Rényi entropy differences can exhibit a dynamical increase that is incompatible with the bounds imposed on asymmetry monotones by the resource-theoretic framework. These examples highlight the importance of distinguishing between quantities that provide useful diagnostics of symmetry breaking and those that additionally quantify it consistently under symmetry-preserving operations. Rather than invalidating the conventional Rényi asymmetries, our results clarify the conditions under which their successful description of symmetry-restoration dynamics admits a resource-theoretic interpretation.

An important methodological outcome of our analysis is that the sandwiched Rényi asymmetry monotones can be accessed through the same charged moments employed in conventional calculations of entanglement asymmetry. The two-parameter replica construction developed here thus provides a bridge between quantum resource theory and the analytical techniques of nonequilibrium many-body physics, including quasiparticle methods, full counting statistics, and space-time duality. This connection opens the possibility of investigating operationally meaningful measures of symmetry breaking in a broad range of quantum many-body systems without abandoning the computational tools that have proved so effective in the study of entanglement and its symmetry resolution.

Several questions naturally arise from our findings. Perhaps the most fundamental is to establish more precisely the conditions under which the correspondence between the two families of Rényi asymmetries holds. Our results suggest that the relevant ingredient is not integrability itself, but rather the possibility of separating, at large scales, an effectively pure component carrying the symmetry-breaking coherence from a symmetric mixed contribution. It would be interesting to understand whether this structure survives in nonintegrable systems, in the presence of diffusive transport, or under more general inhomogeneous dynamics. Conversely, identifying the mechanisms that destroy this effective separation could provide a systematic way to characterize when conventional Rényi asymmetries cease to reproduce the behaviour of genuine monotones. Another natural extension concerns non-Abelian symmetries, for which the structure of charge fluctuations and the associated resource theory are considerably richer, potentially leading to new forms of Rényi-index dependence and quantum Mpemba phenomenology.

Finally, our approach suggests a broader program of comparing analytically accessible probes of nonequilibrium relaxation with quantifiers possessing a precise operational meaning. In this direction, it would be interesting to investigate the distance from the final state, recently proposed in Ref. \cite{Ares_2025_simpler} as a simpler probe of the Mpemba effect, and its relation to resource-theoretic measures of relaxation. Similar questions arise for other quantum resources, such as athermality, recently studied in Ref. \cite{senese2026}, and for the Gaussian asymmetry measure introduced in Ref. \cite{travaglino2026gaussianasymmetrymeasure}, which is monotone under Gaussian symmetry-preserving operations. Understanding whether analogues of the correspondences uncovered here exist for these quantities could reveal which features of nonequilibrium relaxation are genuinely independent of the chosen measure and which encode distinct aspects of the evolving quantum state. More broadly, we hope that the present results will encourage a closer interplay between quantum resource theory and many-body dynamics, providing a framework in which the robustness of phenomena such as symmetry restoration and the quantum Mpemba effect can be assessed beyond any single diagnostic.

\subsection*{Acknowledgments}
We thank Filiberto Ares and Purvaash Panduranghan
 for fruitful discussions. RT and PC acknowledge support from the ERC Advanced Grant MOSE, No. 101199196. MM acknowledges support  from the ERC Starting Grant QUANTHEM, No. 101114881. Views and opinions expressed are however those of the authors only and do not necessarily reflect those of the European Union or the European Research Council Executive Agency. Neither the European Union nor the granting authority can be held responsible for them. \\
\textbf{AI statement}:
AI tools have been used for text polishing and proofreading of the manuscript, including suggestions for improving clarity, grammar, and notation. The authors reviewed and verified all generated content and remain fully responsible for the accuracy, interpretation, and presentation of the results.

\appendix
\section{Bulk vs tail sensitivity in classical Rényi entropies}
 \label{app:bulk_tail} 
In this appendix, we propose a simple example that shows how  classical Rényi entropies $H_\alpha(\{p_q\})= (1-\alpha)^{-1}\log \sum_q p_q^\alpha$ are sensitive to different features of a probability distribution when the index $\alpha$ is varied. 

\begin{figure}[t!]
    \centering
    \includegraphics[width=.8\linewidth]{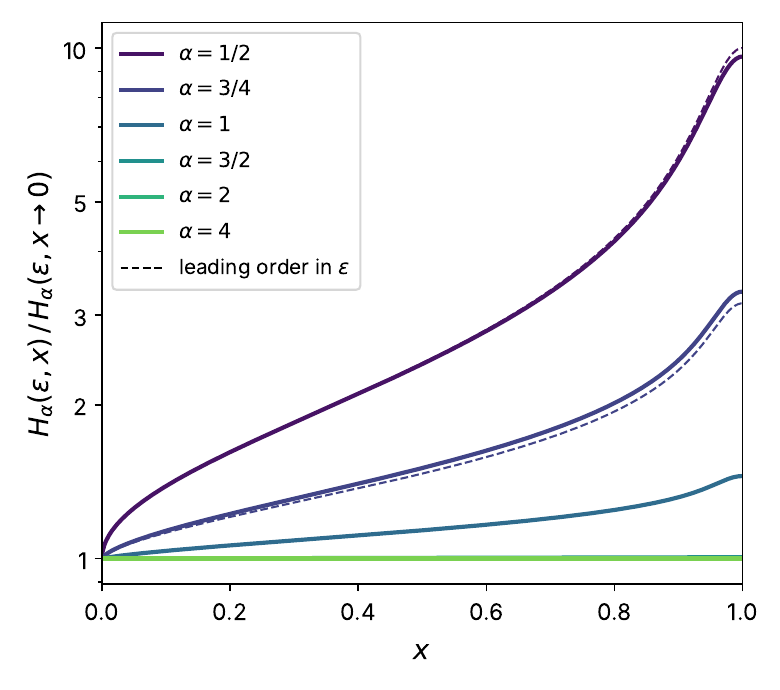}
    \caption{For $N=100$ and $\epsilon=10^{-4}$, we plot the ratio $H_\alpha(\epsilon,x)/H_\alpha(\epsilon,x\to0)$, where the distribution $H_\alpha(\epsilon,x)$ is defined in Eq. \eqref{eq:tail_bulk_renyi_exact} and we choose the geometric tail as specified in the text. As $x\to0$, the tail probability distribution is nonzero only for $q=1$. The plot clearly shows how Rényi entropies with $\alpha >1$ are insensitive to the shape of the tail, as they remain constant throughout the interval $x\in (0,1)$. For $\alpha <1$, the entropies drastically change when the tail is modified. Dashed lines are obtained using the leading order expressions Eqs. \eqref{eq:bulk_leading} (for $\alpha>1$) and \eqref{eq:tail_leading} (for $\alpha<1$).}
    \label{fig:bulk_tail}
\end{figure}

For a small $\epsilon >0$, we construct a probability distribution $\{p_q\}$, for $q=0,\dots,N$, with the following features. The single largest value of $p_q$ is at $q=0$ and constitutes the bulk of the probability distribution: $p_0= 1-\epsilon$. The remaining values are distributed according to a tail probability distribution $\pi_q$: $p_q=\epsilon \pi_q$, $q=1,\dots,N$. This situation, for different shapes of the tail, might model for instance the late stages of the symmetrization process of a pure state, where the charge distribution is strongly peaked around a charge eigenstate. We take the tail distribution $\pi_q$ to be a normalized geometric distribution, $\pi_q=x^{q-1}\frac{1-x}{1-x^N}$, for $x\in (0,1)$. When $x \to 1$, the tail is uniform, as $\pi_q=1/N$ for all $q$, while for $x\to 0$, the tail is only concentrated around $q=1$, as $\pi_1\to 1$, $\pi_{q\ne1}\to 0$. The $\alpha$-Rényi entropy of the distribution $p_q$, which is a function of the parameters $\epsilon$ and $x$, is
\begin{equation}
\label{eq:tail_bulk_renyi_exact}
H_\alpha(\epsilon,x) = \frac{1}{1-\alpha}\log[(1-\epsilon)^\alpha + \epsilon^\alpha G_\alpha(x)],
\end{equation}
with $G_\alpha (x)=\left(\frac{1-x}{1-x^N}\right)^\alpha \frac{1-x^{\alpha N}}{1-x^\alpha}$. Notice that $G_\alpha(x)$ is related to the $\alpha$-Rényi entropy of the tail distribution as $(1-\alpha)H_\alpha(\{\pi_q\})=\log G_\alpha(x)$, hence (since $0\le H_\alpha(\{\pi_q\})\le \log N)$, one has $N^{1-\alpha} \le G_\alpha(x) \le 1$ when $\alpha >1$, and $1 \le G_\alpha(x) \le N^{1-\alpha}$ when $\alpha <1$. The different ranges of $G_\alpha(x)$ control the sensitivity of $H_\alpha$ to the bulk or to the tail distribution. Indeed, expanding in $\epsilon$, for $\alpha>1$ one has
\begin{equation}
\label{eq:bulk_leading}
H_\alpha(\epsilon,x)=\frac{\alpha}{\alpha-1}\epsilon + O(\epsilon^2,\epsilon^\alpha),
\end{equation}
which means that for $\alpha >1$, at leading order $H_\alpha$ is insensitive to the tail probability distribution. On the other hand, for $\alpha <1$ one has $\epsilon^\alpha G_\alpha(x) \gg \epsilon$, and thus 
\begin{align}
H_\alpha(\epsilon,x)&=\frac{1}{1-\alpha}\log[1+\epsilon^\alpha G_\alpha(x)]\nonumber \\ &-\frac{\alpha}{1-\alpha}\frac{\epsilon}{1+\epsilon^\alpha G_\alpha(x)}+O(\epsilon^2),
\end{align}
and in the case $\epsilon^\alpha G_\alpha(x) \ll 1$ (true as long as the parameter $\epsilon$ is chosen $\epsilon \ll N^{-(1-\alpha)/\alpha}$) then, more explicitly
\begin{equation}
\label{eq:tail_leading}
H_\alpha(\epsilon,x) = \frac{G_\alpha(x)\epsilon^\alpha}{1-\alpha} + O(\epsilon,\epsilon^{2\alpha}),
\end{equation}
which shows that, for $\alpha <1$, $H_\alpha$ depends on the tail of the probability distribution at leading order. These features, discussed in the case of a geometric tail, are independent of the specific shape $G_\alpha$ of the tail probability distribution. The different sensitivities of the Rényi entropies $H_\alpha(\{p_q\})$ for different values of the Rényi index $\alpha$ are summarized in Fig. \ref{fig:bulk_tail}.

\section{Interacting integrable generalisation}
\label{app:interacting}
In this appendix, we comment on the generalisation of the results of section \ref{sec:replicatrick} to interacting integrable quenches. Without loss of generality, we focus on a system with a single particle species; for more general situations, it is enough to sum over the different bound states which determine the spectrum of the model. In general, integrable models modify free theories through the introduction of a scattering kernel $T(\mu,\mu')$ which takes into account the allowed (elastic) scatterings. Typically they are treated by introducing a rapidity variable $\mu$ which conveniently parametrizes the energy-momentum dispersion relation, $(\varepsilon(\mu),p(\mu))$. Integrability then guarantees that the effect of interactions can be in general reabsorbed through an exact renormalisation of physical quantities, including charges $q \to q^{dr}$, particle densitites $\frac{1}{2\pi} \to \rho_t$ and  propagation velocities $v\to v^{dr}$, where the dressing operation amounts to a relatively simple integral equation involving the scattering kernel \cite{bertini2016transport,castroalvaredo2016emergent}.

The charged moments for this scenario were derived in \cite{bertini2024dynamics,travaglino2026space}, as 
\begin{equation}
    \log Z_n^A(\lambda_1, \dots \lambda_n) = \ell \log \Lambda_{\{\lambda_i\}} + 2t \,s_{\{\lambda_i\}},
\end{equation}
where the two terms in the right-hand side account for the initial state contribution and the slope of the linear growth of the charged moments. Using $\lambda=\sum_j \lambda_j$, these are expressed as 
\begin{equation}
\label{eq:interacting_charged_moments}
\begin{split}
     \log \Lambda_{\{\lambda_i\}} &=  \sum_{j=1}^n \int \frac{d\mu}{4\pi} p'(\mu) \log\left(1-\vartheta(\mu) + \frac{\vartheta(\mu)}{x_{2\lambda_j}} \right),\\  s_{\{\lambda_i\}}&= \int \frac{d\mu}{2\pi} |\varepsilon'(\mu)| \log\left(\prod_j (1-\vartheta_j(\mu)) + \frac{\prod_j \vartheta_j(\mu)}{y_{\lambda}^{\text{sign}(\varepsilon'(\mu))}}\right),  
\end{split}
\end{equation}
where $\vartheta_j$ is obtained from a bipartitioning protocol induced by the action of $e^{\lambda_j Q_A}$ on the initial state, see \cite{bertini2024dynamics,horvath2025a,horvath2025counting}, and we have introduced the two auxiliary quantities 
\begin{equation}
    \begin{cases}
        \log x_{2\lambda_j} = - 2\lambda_j q + \int \frac{d\mu'}{2\pi} T(\mu',\mu) \log\left(1-\vartheta(\mu) + \frac{\vartheta(\mu)}{x_{2\lambda_j}} \right) \\
         \log y_{\lambda} = - \lambda q + \int \frac{d\mu}{2\pi} T(\mu',\mu) \log\left(\prod_j (1-\vartheta_j(\mu)) + \frac{\prod_j \vartheta_j(\mu)}{y_{\lambda}^{\text{sign}(\varepsilon'(\mu))}}\right)  .
    \end{cases}
\end{equation}
Such charged moments can then be evaluated with the parameters \eqref{lambdalist} which arise in the replica trick. Contrarily to the free case, the function $s_{\lambda_i}$ couples all the parameters nontrivially, and therefore the factor $n$ (one of the two numbers of replicas, see Eq. \eqref{eq:physical_limit}) cannot be extracted exactly as was done in the main text. However, the same features still emerge at the saddle-point level. The first simplification arises from the consideration that \eqref{lambdalist} involves a grouping of parameters $\lambda_{j,(k+1,k)}$ such that $\sum_k \lambda_{j,(k+1,k)} =0$, which implies also $\sum_{k,j} \lambda_{j,(k+1,k)}=0$; furthermore, as observed in \cite{bertini2024dynamics}, the expressions \eqref{eq:interacting_charged_moments} are invariant under all the permutations of the parameters $\lambda_j$. This implies that the Hessian $\mathbb H$ determining the quadratic term does not couple the blocks at different $j$, which therefore contribute additively. The only coupling remains between the different values of $k$, which involve terms $\lambda_{j,(k+1,k)}=\lambda_{j,k+1}-\lambda_{j,k}$. To see this, note that permutation symmetry implies that 
\begin{equation}
    \mathbb H_{(j,k),(j',k')} =  \mathbb H_{\rm d} \delta_{j,j'}\delta_{k,k'} +  \mathbb H_{\rm nd } (1-\delta_{j,j'}\delta_{k,k'}),
\end{equation}
which in turn implies that the quadratic order of the exponential obeys 
\begin{equation}
    \sum_{j,j'}\sum_{k,k'} \mathbb H_{(j,k),(j',k')} \lambda_{j,(k+1,k)} \lambda_{j',(k'+1,k')} = \sum_{j,k} ( \mathbb H_{\rm d} -\mathbb H_{\rm nd } ) \lambda_{j,(k+1,k)}^2 +\mathbb H_{\rm nd }\left(\sum_{j,k}  \lambda_{j,(k+1,k)} \right)^2,
    \end{equation}
    where by the above discussion the second term, which contains the coupling between different $j$ sectors, vanishes. Since there is no coupling at quadratic order between the different $j$ blocks, the saddle point solution will be the solution for a single block repeated $n$ times, giving 
\begin{equation}
     S^{(m,n)} \simeq n \log \int \prod_{k=1}^m \frac{d\lambda_k}{2\pi}  \exp\left\{\sum_{k=0}^{m}\mathcal{F}(\lambda_{k+1,k})\right\}.
\end{equation}
In the above expression, $\mathcal{F}$ is the quadratic contribution of the exponent of the charged moments, which is identical to the one appearing in \cite{bertini2024dynamics} up to the extra boundary parameters discussed in the main text. Hence the discussion can be concluded proceeding by analogy with the cited reference:
the saddle point values are $\lambda_{k}=0 \,\, \forall j,k$, and the derivatives satisfy $\partial_{\lambda_{q}} \partial_{\lambda_{k}}Z|_{\{\lambda=0\}} = \partial^2_{\lambda_{k}} Z|_{\{\lambda=0\}}$, leading to the saddle point result in the physical limit
\begin{equation}
      S^{(\alpha)}(\rho_A||\rho_{A,Q})  \simeq \frac{1}{2}\log\left(\pi \alpha^{\frac{\alpha}{\alpha-1}}\int d\mu \, \ell_{\mu}(x,t) \chi(\mu)  \right), \hspace{0.3cm}
     \chi(\mu) = \rho_t(\mu)\, \vartheta(\mu) (1-\vartheta(\mu)) (q^{dr})^2,
\end{equation} 
which, compared to the result of \cite{bertini2024dynamics}, confirms the $\alpha \leftrightarrow 1/\alpha$ structure.

\bibliographystyle{ytphys}
\bibliography{bibliography.bib}
\end{document}